# Probing dynamics and pinning of single vortices in superconductors at nanometer scales

L. Embon[1,*], Y. Anahory[1,*], A. Suhov[1], D. Halbertal[1], J. Cuppens[1], A. Yakovenko[1], A. Uri[1], Y. Myasoedov[1], M.L. Rappaport[1], M.E. Huber[2], A. Gurevich[3] and E. Zeldov[1]

[1]Department of Condensed Matter Physics, Weizmann Institute of Science, Rehovot, 7610001, Israel
[2]Department of Physics, University of Colorado Denver, Denver, 80217, USA
[3]Department of Physics, Old Dominion University, Norfolk, VA 23529-0116, USA
[*]These authors contributed equally to this work
Corresponding authors: lior.embon@weizmann.ac.il, yonathan.anahory@weizmann.ac.il

**The dynamics of quantized magnetic vortices and their pinning by materials defects determine electromagnetic properties of superconductors, particularly their ability to carry non-dissipative currents. Despite recent advances in the understanding of the complex physics of vortex matter, the behavior of vortices driven by current through a multi-scale potential of the actual materials defects is still not well understood, mostly due to the scarcity of appropriate experimental tools capable of tracing vortex trajectories on nanometer scales. Using a novel scanning superconducting quantum interference microscope we report here an investigation of controlled dynamics of vortices in lead films with sub-Angstrom spatial resolution and unprecedented sensitivity. We measured, for the first time, the fundamental dependence of the elementary pinning force of multiple defects on the vortex displacement, revealing a far more complex behavior than has previously been recognized, including striking spring softening and broken-spring depinning, as well as spontaneous hysteretic switching between cellular vortex trajectories. Our results indicate the importance of thermal fluctuations even at 4.2 K and of the vital role of ripples in the pinning potential, giving new insights into the mechanisms of magnetic relaxation and electromagnetic response of superconductors.**

The ability to carry non-dissipative electric currents in strong magnetic fields is one of the fundamental features of type-II superconductors crucial for many applications [1-10]. The current, however, exerts a transverse Lorentz force on vortices, leading to their dissipative motion and to finite resistance unless materials defects immobilize (pin) vortices at current densities below some critical value $J_c$. Pinning potential wells $U(r)$ produced by defects are the key building blocks that determine collective pinning phenomena, in which a flexible vortex line is pinned by multiple defects. The global electromagnetic response of the vortex matter is thus governed by a complex interplay of individual pinning centers, interaction between vortices, and thermal fluctuations [11]. Recent advances in materials science have enabled several groups to controllably produce nanostructures of nonsuperconducting precipitates to optimize the pinning of vortices in superconductors and to achieve $J_c$ up to 10-30% of the fundamental depairing current density $J_d$ at which the current breaks Cooper pairs [2-10]. In artificially-engineered pinning structures, the shape of $U(r)$ can also be made asymmetric to produce the intriguing vortex ratchet and rectification phenomena [12-16]. Local studies of individual vortices using scanning probe techniques have revealed phase transitions and depinning of vortex lines and the collective motion of the vortex lattice [17-30]. Even so, the intrinsic structure of a single pinning potential well $U(r)$, which determines the fundamental interaction of vortices with pinning centers, has not yet been measured directly.

Difficulties with the measurement of $U(r)$ arise from the necessity of deconvoluting the effect of multiple pinning defects along the elastic vortex line, and from the lack of experimental techniques for extracting $U(r)$ on the scale of the superconducting coherence length $\xi$ (ranging from 2 to 100 nm for different materials) that quantifies the size of the Cooper pair. To avoid the complexity of the situation in which the vortex line behaves like an elastic string pinned by multiple defects [21], the vortex length should to be of the order of $\xi$. This situation can be achieved in thin superconducting films of thickness $d \cong \xi$ in a perpendicular magnetic field. Moreover, in order to simplify the interpretation of experimental data, it is desirable to choose a film in which the three basic length scales of the system are comparable, $\xi \cong \lambda \cong d$, where $\lambda$ is the bulk magnetic penetration depth. We have therefore chosen to study Pb thin films ($T_c$ = 7.2 K) with $d = 75$ nm, $\xi(4.2\ \text{K}) = 46.4$ nm, and $\lambda(4.2\ \text{K}) \cong 90$ nm. In this case, the vortex can be regarded as a particle in a random two-dimensional (2D) potential landscape of pinning defects.

We employed a novel nanoscale superconducting quantum interference device (SQUID) that resides on the apex of a sharp tip [31,32] to trace vortex trajectories. A highly sensitive SQUID-on-tip (SOT) with a diameter of 177 nm was integrated into a scanning probe microscope operating at 4.2 K [32]. Figure 1a shows a scanning magnetic image of a Pb film patterned into an 8 μm wide microbridge in which a single vortex was trapped upon field cooling at about 0.1 mT. A small ac current $I_{ac}$ is applied along the bridge ($y$ direction) resulting in a Lorentz force on the vortex $F_{ac} = \Phi_0 J_{ac}$ along the perpendicular $x$ axis, where $J_{ac}$ is the sheet current density and $\Phi_0$ = 2.07×10$^{-15}$ Wb is the magnetic flux quantum. The weak $F_{ac}$ results in a small oscillation of the vortex around its equilibrium position with typical amplitude of 1 nm, much smaller than $\xi$. The scanning SOT microscope simultaneously measures the distributions of the dc magnetic field $B_{dc}(x, y)$ and of the ac field $B_{ac}(x, y)$ at the driving frequency measured by a lock-in amplifier (Figures 1a and 1b). For small displacements of the vortex, the two fields are related by

$$B_{ac} = -x_{ac}\frac{\partial B_{dc}}{\partial x} - y_{ac}\frac{\partial B_{dc}}{\partial y}. \qquad (1)$$

Using this relation, we measured the ac displacements $x_{ac}$ and $y_{ac}$ of the vortex along the x and y axes as outlined in Figures 1c to 1g. The very high sensitivity of our SOT allows us to measure sub-atomically-small displacements, down to 10 pm [Figure S7]. By adding a dc current $I_{dc}$, a driving force $F_{dc} = \Phi_0 J_{dc}$ is exerted that tilts the potential and shifts the vortex equilibrium position. By changing $F_{dc}$ in small steps, we are able to reconstruct the full vortex ac response $x_{ac}(F_{dc})$ and $y_{ac}(F_{dc})$ along the potential well. If the vortex oscillates within a potential well without hopping to neighboring wells, dissipation is negligible and $x_{ac}$ and $y_{ac}$ are in-phase with $I_{ac}$.

We first analyze the expected response of the vortex trapped in a generic potential well [11,33-35] $U(r) = -U_0/(1 + (r/\xi)^2)$ produced by a materials defect of size $a \ll \xi$, where $r$ is the radial distance from the defect (Figure 2d). Depending on the nature and the size of the defect, the pinning energy $U_0$ is a fraction of the maximum superconducting condensation energy of the vortex core

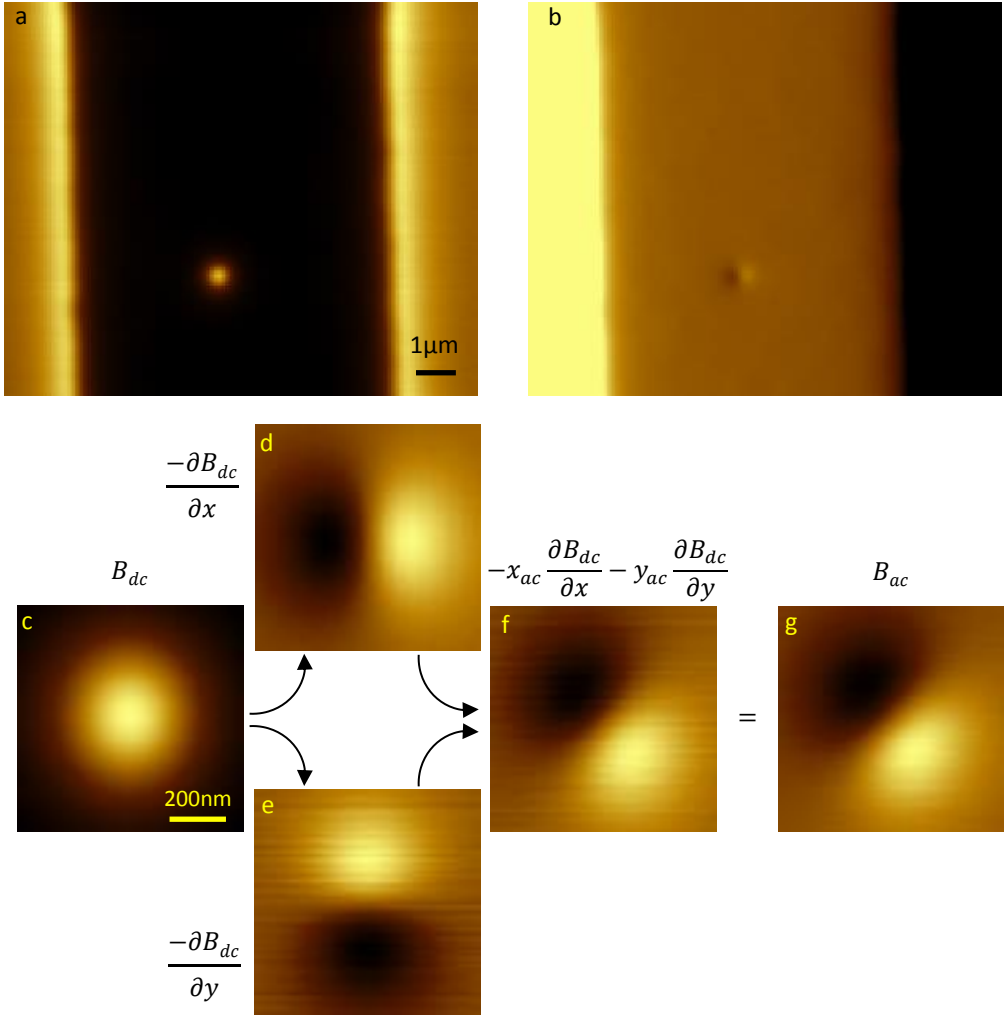

**Figure 1. Magnetic imaging of a single vortex. a.** Scanning SQUID-on-tip image of $B_{dc}(x,y)$ showing a single vortex (bright) in a thin Pb film patterned into an 8 μm wide microbridge at $T = 4.2$ K. The microbridge is in the Meissner state (dark), and the enhanced field outside the edges (bright) is due to the screening of the applied field of 0.3 mT. **b.** Scanning image of $B_{ac}(x,y)$ acquired simultaneously with (**a**) showing the vortex response to an ac current of $I_{ac} = 0.94$ mA peak-to-peak (ptp) at 13.3 kHz applied to the microbridge. The Meissner response is visible along the microbridge ($B_{ac} = 0$, light brown) with positive (negative) $B_{ac}$ outside the left (right) edge due to the field self-induced by $I_{ac}$. **c.** A zoomed-in image of the measured $B_{dc}(x,y)$ of a vortex. **d - f.** Numerically derived $\partial B_{dc}/\partial x$, $\partial B_{dc}/\partial y$ and $-x_{ac}\partial B_{dc}/\partial x - y_{ac}\partial B_{dc}/\partial y$ with $x_{ac} = 1.6$ nm and $y_{ac} = -1.9$ nm values obtained by a fit to (**g**). **g.** Experimentally measured $B_{ac}(x,y)$ of the vortex driven by $I_{ac}$ acquired simultaneously with (**c**).

$U_p \cong H_c^2 \xi^2 d = 28.3$ eV [11], where $d = 75$ nm is the film thickness, $H_c(4.2 \text{ K}) = 530$ Oe is the thermodynamic critical field of Pb, and $\xi(4.2 \text{ K}) = 46.4$ nm is derived from the measurement of the upper critical field $H_{c2} = \Phi_0/2\pi\xi^2$ (Figure S5). In the absence of current, the vortex is located at the minimum of $U(r)$ at $x_m = 0$. A dc driving force $F_{dc} = \Phi_0 J_{dc}$ tilts the potential, $U_f(x) = U(x) - F_{dc}x$, displacing the vortex to a new minimum at $x_m(F_{dc})$, where the driving force is balanced by the restoring force, $F_r = -F_{dc} = -dU/dx|_{x_m}$ (Figure 2c). In our experiment, we measure the ac displacement of the vortex $x_{ac}(F_{dc}) = -F_{ac}/(\partial F_r/\partial x|_{x_m}) = F_{ac}/(\partial^2 U/\partial x^2|_{x_m})$ which is inversely proportional to the pinning spring constant, $k = \partial^2 U/\partial x^2$. Figure 2a shows that the spring constant $k(x)$ is the stiffest in the center of the well and gradually softens towards the inflection points of $U(x)$ at $x_i = \pm\xi/\sqrt{3}$ so that $x_{ac}(F_{dc})$ progressively increases until the vortex hops out of the well.

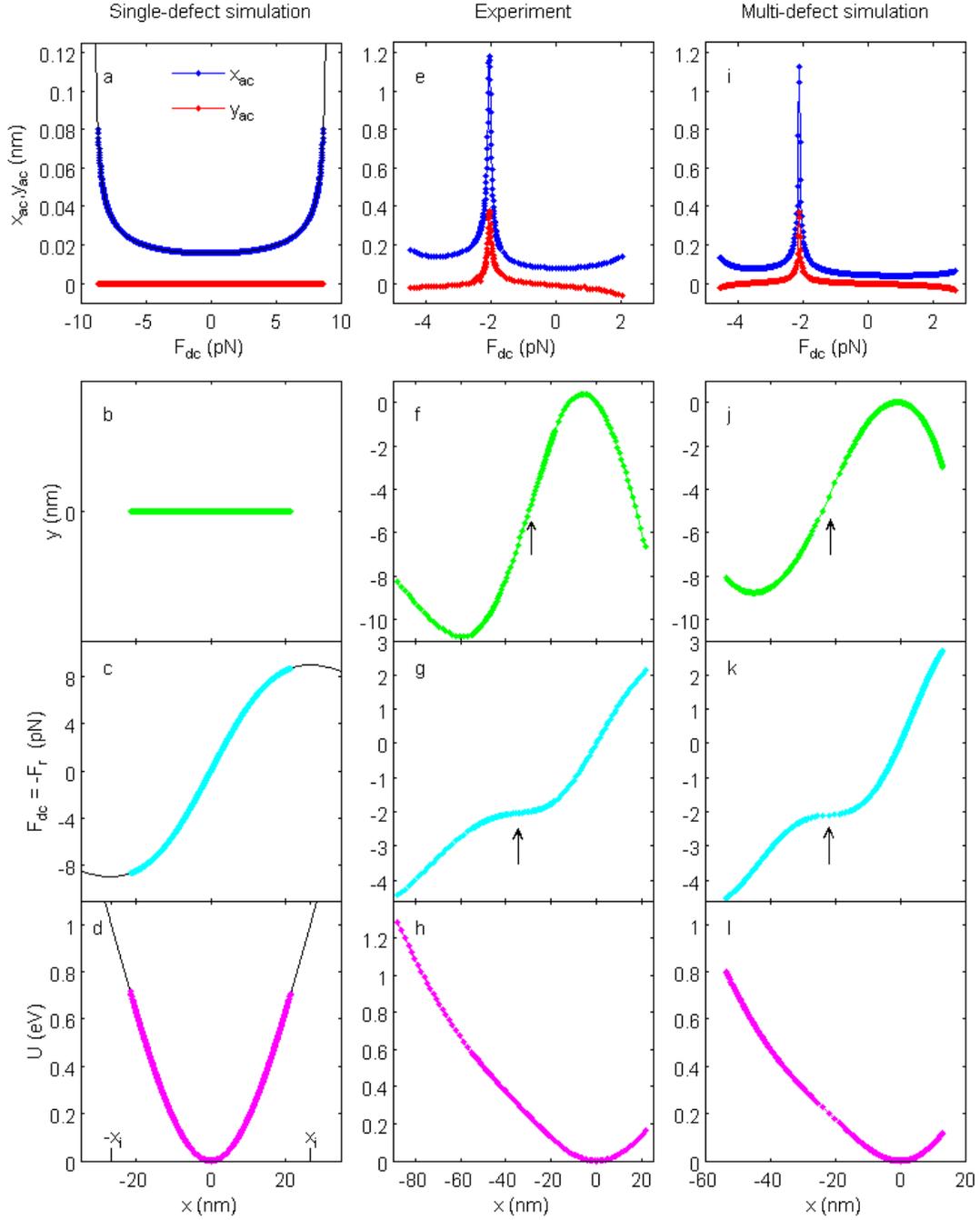

**Figure 2. Vortex response and the structure of potential well 2, and comparison to simulations.**
**a-d**, Calculated vortex response to $F_{ac} = 8.89$ fN in a potential well $U(r) = -U_0/(1+(r/\xi)^2)$ due to a single defect with $U_0 = 4$ eV: **a**, vortex ac displacement $x_{ac}$ and $y_{ac}$; **b**, vortex trajectory; **c**, restoring force; and **d**, the potential. The data points end where the activation barrier $\Delta U = 34 k_B T$ (see S9). **e-h**, Measured vortex response in well 2: **e**, displacements $x_{ac}$ and $y_{ac}$ in response to $F_{ac} = 8.89$ fN, displaying a large softening peak in the middle of the well; **f**, vortex trajectory with an 'S' shape; **g**, restoring force showing a pronounced inflection point; and **h**, the potential well. **i-l**, Calculated vortex response in a potential well due to a cluster of four defects (see Figure 5a for locations of defects): **i**, ac vortex displacement; **j**, vortex trajectory; **k**, restoring force; and **l**, the potential. The data points end where $\Delta U = 34 k_B T$.

By sweeping the driving force $F_{dc}$ and integrating over $x_{ac}$, the $x$ position of the vortex $x_m(F_{dc}) = \int (dx/dF)\, dF_{dc} = \int (x_{ac}/F_{ac})\, dF_{dc}$ and the shape of the restoring force $F_r(x) = -F_{dc}(x_m)$ are obtained. Combining these results with the corresponding integration over $y_{ac}$ yields the full vortex trajectory $y_m(F_{dc})$ vs. $x_m(F_{dc})$ within the well. In an isotropic potential, $y_{ac} = 0$ and the vortex trajectory is a straight line (Figure 2b).

Figure 2e shows the ac displacements $x_{ac}(F_{dc})$ and $y_{ac}(F_{dc})$ measured for one of the potential wells (well 2, see Figure 3). Following the above procedure, we derive the vortex trajectory within the well (Figure 2f), the restoring force (Figure 2g), and the single-well pinning potential $U(x) = \int F_{dc}(x)dx$ (Figure 2h). Figure 3 shows a compilation of similar results for different potential wells. These data reveal the following features of vortex response that are strikingly different from the expected behavior shown in Figures 2a-d. (*i*) *Spring softening in the middle of the well*. In contrast to Figure 2a, which shows the highest stiffness in the center of the well, Figure 2e reveals a large and sharp peak in $x_{ac}$, which implies a small $k$ and an inflection point in the restoring force in the central region of the well, shown by the arrow in Figure 2g. This intriguing feature turned out to be ubiquitous, and various degrees of softening in the middle of the wells were found in the majority of pinning sites as illustrated by Figures 3b and S8d. (*ii*) *Broken-spring phenomenon*. The ac response $x_{ac}(F_{dc})$ is expected to increase progressively towards the edges of the well and diverge at the inflection point of $U(x)$ due to the vanishing spring constant $k(x)$ at the maximum restoring force (Figures 2a and 2c). Surprisingly, Figure 2e shows only a small increase in $x_{ac}$ towards the well edges, where the restoring force (Figure 2g) exhibits hardly any rounding at its maximum values. This response resembles the abrupt breaking of an elastic spring. (*iii*) *Anisotropy*. In an isotropic well, the vortex moves only in the direction of the driving force (Figures 2a and 2b). However, Figure 2e shows that the vortex also has a significant transverse ac displacement $y_{ac}$ and a substantial $y$ component in the trajectory (Figure 2f). Analysis of different wells has shown that the vortex may move at angles as large as 77° with respect to the direction of the driving force (see Figures 3 and S8). (*iv*) *Asymmetry and internal structure*. Figures 3c and S8e show that most of the potential wells are asymmetric with respect to the positive and negative driving force and exhibit significant deviations from the model function $U(r) = -U_0/(1 + (r/\xi)^2)$. In Figure 2h, for example, the well is nearly parabolic at the bottom but has a linear intermediate section that again becomes parabolic at larger negative displacements. Even though the wells appear to be smooth on the scale of $\xi$, they have an internal structure that results in nontrivial shapes of the restoring forces and intricate trajectories of vortices in the individual wells (Figures 3a-b and S8c-d). Also, a point defect results in a well of width $\sim 2\xi/\sqrt{3} \cong 53$ nm between the inflection points, while many of the observed wells are significantly wider, reaching 110 nm (Figures 3c and S8e). (*v*) *Correlation between softening and the inflection point in the trajectory*. We also observed an intriguing correlation between the position of the softening point of the potential, which is the inflection point in the restoring force, and the position of the inflection point in the trajectory, as marked by arrows in Figures 2f and 2g. This correlation gives an important clue to the internal structure of the pinning potential and the dynamics of vortices, as described below.

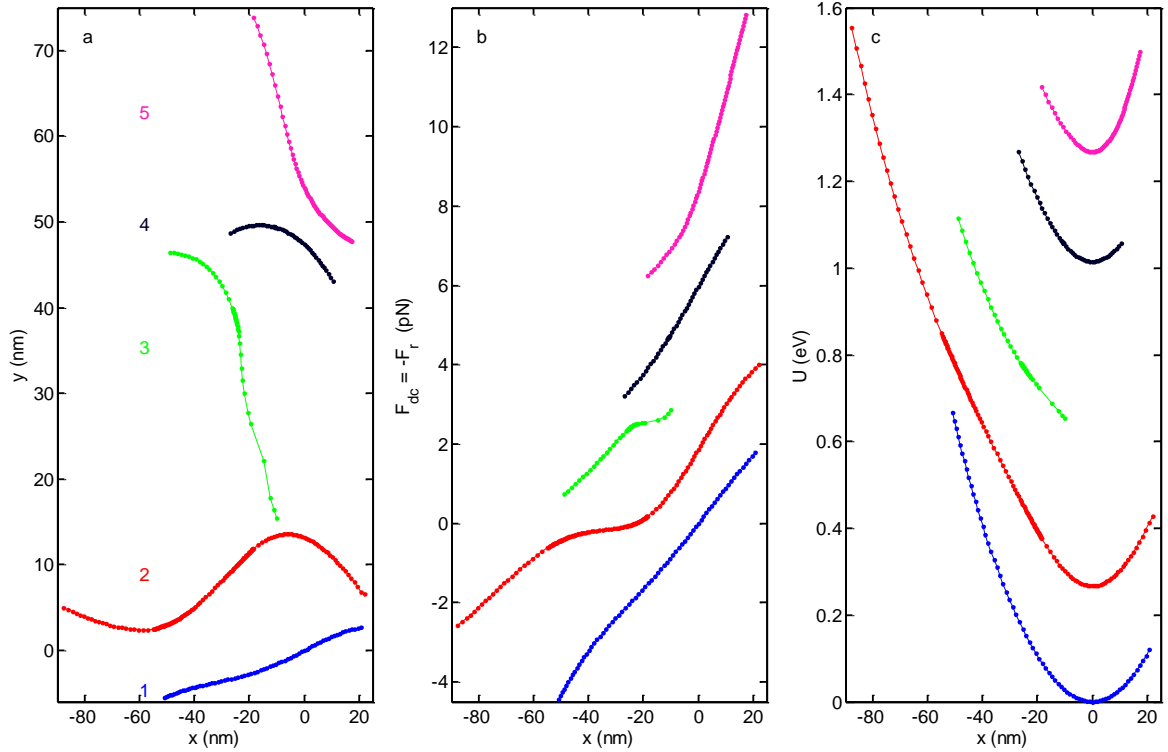

**Figure 3. Comparison of different potential wells. a**, Vortex trajectories in wells 1 to 5, shifted vertically for clarity. $x = 0$ corresponds to the rest position of the vortex at $F_{dc} = 0$, except for the metastable well 3 that does not exist at $F_{dc} = 0$ (see Figure 4b). All the wells display nontrivial internal structure. **b**, The restoring force $F_{dc} = -F_r(x)$ of the different wells shifted vertically for clarity. $x = 0$ corresponds to $F_{dc} = 0$ for each well except well 3. **c**, The structure of the potential of the different wells shifted for clarity.

We now examine vortex dynamics on larger scales that include several pinning wells. At each value of $F_{dc}$, we acquire a full image of $B_{dc}(x, y)$ and $B_{ac}(x, y)$ at a constant $F_{ac}$, compiling a movie of the ac response as $F_{dc}$ is swept back and forth (See S6). Figures 4e-j show several frames of $B_{ac}(x, y)$ from one of the movies. As $F_{dc}$ is changed, $B_{ac}(x, y)$ shows significant variations in the intensity and the orientation of the dipole-like signal, reflecting the changes in both amplitude and direction of the ac displacement of the vortex within a single well. If $F_{dc}$ exceeds the maximum restoring force of a well, the vortex jumps to a different well, manifesting as an instantaneous displacement of the dipole in the movie. By recording the ac displacements in the wells and the jumps between the wells, a full map of closed loops of vortex trajectories was obtained, as shown in Figure 4a (and an additional example in Figure S8a). The corresponding plot of the restoring force is shown in Figure 4b (and S8b). Figure 4a shows that the vortex resides in well 1 at large negative values of $F_{dc}$. As the applied force exceeds the maximum restoring force, the vortex may jump all the way to well 5, where it stays up to our maximum $F_{dc} = 4.44$ pN. As $F_{dc}$ is decreased, the vortex undergoes a sequence of jumps to wells 4, 3, 2, and eventually back to 1. While the trajectory of the vortex within a well is reversible, the transition between the wells is hysteretic. Thus, to explore all possible intermediate trajectories and transitions, sub-loop sweeps of $F_{dc}$ were performed. Analysis of the data from repeated sweeps leads to the following additional conclusions about the interaction of vortices with pinning wells: (*vi*) *Hopping distance and direction*. While the size of the individual wells is typically (1-2)$\xi$, the hopping distance between the wells is usually significantly larger, up to 20$\xi$. The vortex jumps between wells often have large transverse components with respect to the direction of the Lorentz force, as seen in Figure 4a (and S8a). The hopping between wells 3 and 4, for example, occurs at the angle ~ 70° with respect to $F_{dc}$. (*vii*) *Metastable wells*. Some minima in the pinning potential only appear at a finite dc Lorentz force. For example, well 3 in Figure 4b exists only at negative $F_{dc}$ and well 6 exists only at positive $F_{dc}$. Moreover, Fig. 4b shows that the vortex jumps from well 3 to well 4 in the positive $x$ direction even though the driving force is negative (green dashed line); that is, the vortex moves against the Lorentz force. (viii) *Nondeterministic hopping*. Upon repeating the loops several times, we find that vortex jumps between the wells are not deterministic. For example, the force $F_{dc}$ at which the vortex jumps from well 5 to well 4 is quite different for different dc current sweeps (dashed magenta lines in Fig. 4b). In addition, the vortex occasionally hops from well 5 directly to well 3 instead of well 4. Similarly, from well 4 the vortex may jump to wells 3, 2, or 1, while from well 3 it may jump to wells 2 or 1.

The above results clearly indicate that the observed vortex dynamics cannot be described by a sparse distribution of pinning centers that are well separated from each other. Our detailed numerical investigation of random disorder shows that the vortex response changes drastically if the potential wells of individual defects overlap. For instance, two small defects separated by a distance $l > 2\xi/\sqrt{3}$ form two distinct potential wells separated by a barrier (Figure S10a). However, at smaller separations, $\xi/2 \lesssim l < 2\xi/\sqrt{3}$, an interesting situation occurs in which the two defects form a single potential well, but $U(x)$ develops an intrinsic softening and an inflection point in the restoring force leading to a peak in $x_{ac}$ in the center of the well, as shown in Figures. S9a-c. For two identical defects, the softening occurs exactly in the center of the well where the attractive forces of the two defects balance each other, leading to a U-shaped potential well. This symmetric configuration is readily perturbed if the defects have different $U_0$ or if other defects are nearby.

By analyzing a cluster of four defects, we can reproduce the results of our SOT vortex microscopy with startling agreement between the experimental and numerical results shown in Figure 2i-l: 1) The vortex displacement has a sharp spring softening peak in the middle of the well (Figure 2i) and a corresponding inflection point in the restoring force (Figure 2k). 2) The potential structure is substantially wider than $\xi$. 3) There is a significant anisotropy of the vortex response (Figure 2j) and asymmetry between positive and negative drives (Figure 2l). This anisotropic vortex response occurs here in a cluster of *isotropic* pinning defects, unlike the anisotropic depinning in a network of planar defects such as grain boundaries in polycrystals [34]. 4) The trajectory has an S shape with an inflection point (arrow in Figure 2j) that clearly matches the location of the inflection point in the restoring force (arrow in Figure 2k), as we indeed observe experimentally.

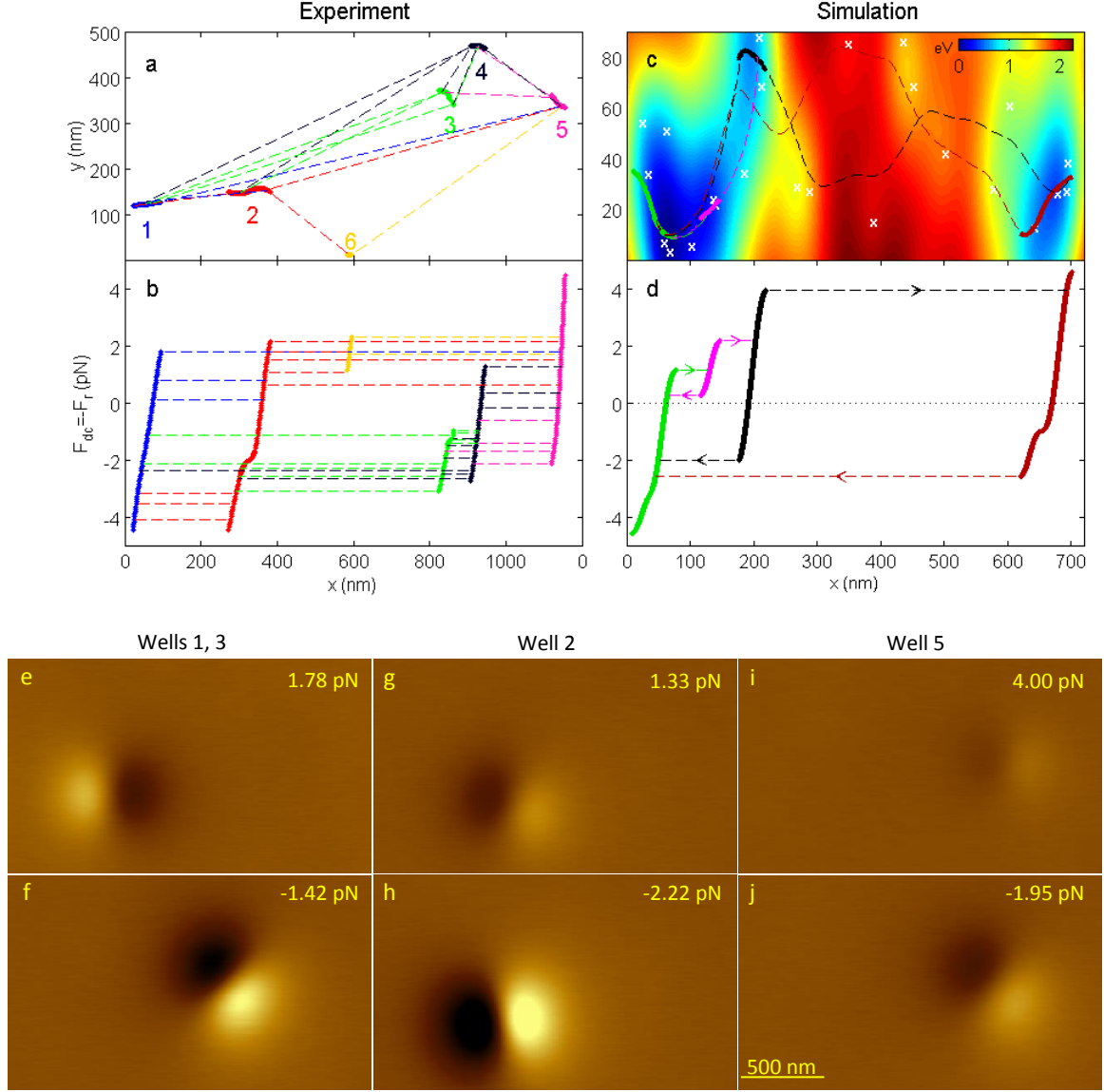

**Figure 4. Vortex hopping between wells. a**, Vortex trajectories in wells 1 to 6 (solid symbols) and hopping events between the wells shown schematically by dashed lines with a color matching the original well. The hopping between the wells results in hysteric closed-loop trajectories that vary upon repeated cycles of the full loop and of the sub-loops. **b**, The restoring force of the wells $F_{dc} = -F_r(x)$ with the hopping events (dashed lines with a color matching the original well). The vortex jumps to the right (left) when a positive (negative) applied force exceeds the restoring force. The vortex stops at a position where the restoring force of the new well equals the applied force (dashed lines are horizontal). Well 6 exists only at positive forces and well 3 only at negative forces. The jumps to the right from well 3 to well 4 occur against the direction of the driving force when the value of the applied force drops below the minimum restoring force of well 3. **c**, 2D potential due to a random distribution of defects, marked by ×, of equal $U_0 = 0.66$ eV and average density of 200 µm$^{-2}$. The stationary trajectories upon tilting the potential are shown by the solid lines and the dynamic escape trajectories by dashed lines with colors matching the source well. **e-j**, Selected $B_{ac}(x, y)$ images from a movie (S6) of the ac vortex response to $F_{ac} = 93.1$ fN ptp in different wells at the indicated values of $F_{dc}$. The gray scale spans 20 µT in all images.

The analysis of this multi-defect cluster also provides an important insight into the observed broken-spring effect. At zero driving force, the four close defects form a single potential well with its minimum at the origin. The blue solid line in Figure 5b shows the resulting potential $U(x)$ along the line $y = 0$, parallel to the driving force. The corresponding restoring force $\partial U/\partial x$ (blue line in Figure 5c), is smooth and flattens out at its maximum and minimum values, similar to that of a single defect (Figure 2c). However, as the potential is tilted by the driving force, the vortex moves along an intricate 2D trajectory shown in Figure 5a. In particular, for a small positive $F_{dc}$, only one potential minimum (magenta dot in Figure 5e) and one saddle point (black dot) are present. At $F_{dc} = 1.04$ pN, however, a new metastable well appears with a new minimum and saddle point (light green and yellow dots in Figures 5f and 5g). This additional well produces a sharp ripple in the projection of the potential $U(x)$, and of the corresponding restoring force, onto the direction of the driving force (Figures 5b and 5c). Thus, in contrast to the common perception that the pinning potential of multiple defects is smooth on the scale of $\xi$, a multi-defect potential along the projection of the intricate vortex trajectory onto the direction of the Lorentz force can have sharp ripples on length scales substantially shorter than $\xi$. Because of these ripples, the restoring force terminates quite abruptly as $F_{dc}$ approaches a narrow region near the end points of the magenta line in Figure. 5c, giving rise to the broken-spring response. Interestingly, Figures 5a and 5c show that, upon exiting the central (magenta) well to the right, the vortex hops into the green metastable well, while, upon exiting to the left, the vortex escapes without passing through the red metastable well. Thus, the effects of the ripples persist even if the vortex does not actually hop into the metastable wells.

Broken-spring behavior is usually associated with the so-called pin-breaking mechanism that results from bending distortions of a long vortex trapped by a strong pinning center [36-39]. In our case, however, bending distortions of the vortex are suppressed because the thickness of our film is smaller than the diameter of the non-superconducting core, $2^{3/2}\xi = 131$ nm. A pinhole of radius $a > \xi$ with sharp boundaries can also result in a hysteretic pinning potential $U(x, y)$ with no reversible quadratic part at small displacements [40], which is inconsistent with our SOT data shown in Figure 2. Our numerical simulations of the Ginzburg-Landau equations (see S13) for the case of strong pinning (due to $T_c(r)$ depression in a region of radius $a \sim \xi$ and smooth recovery over the length $\sim \xi$ at $r > a$) gave a potential well $U(r)$ similar to that shown in Figure 2d, including the softening of the spring constant at the inflection point (Figure S11). Thus, our model of potential ripples in a 2D random potential formed by an array of overlapping single wells $U(r)$ is applicable to both weak and strong pinning and it presents a new mechanism for spring-breaking that describes our experimental data surprisingly well.

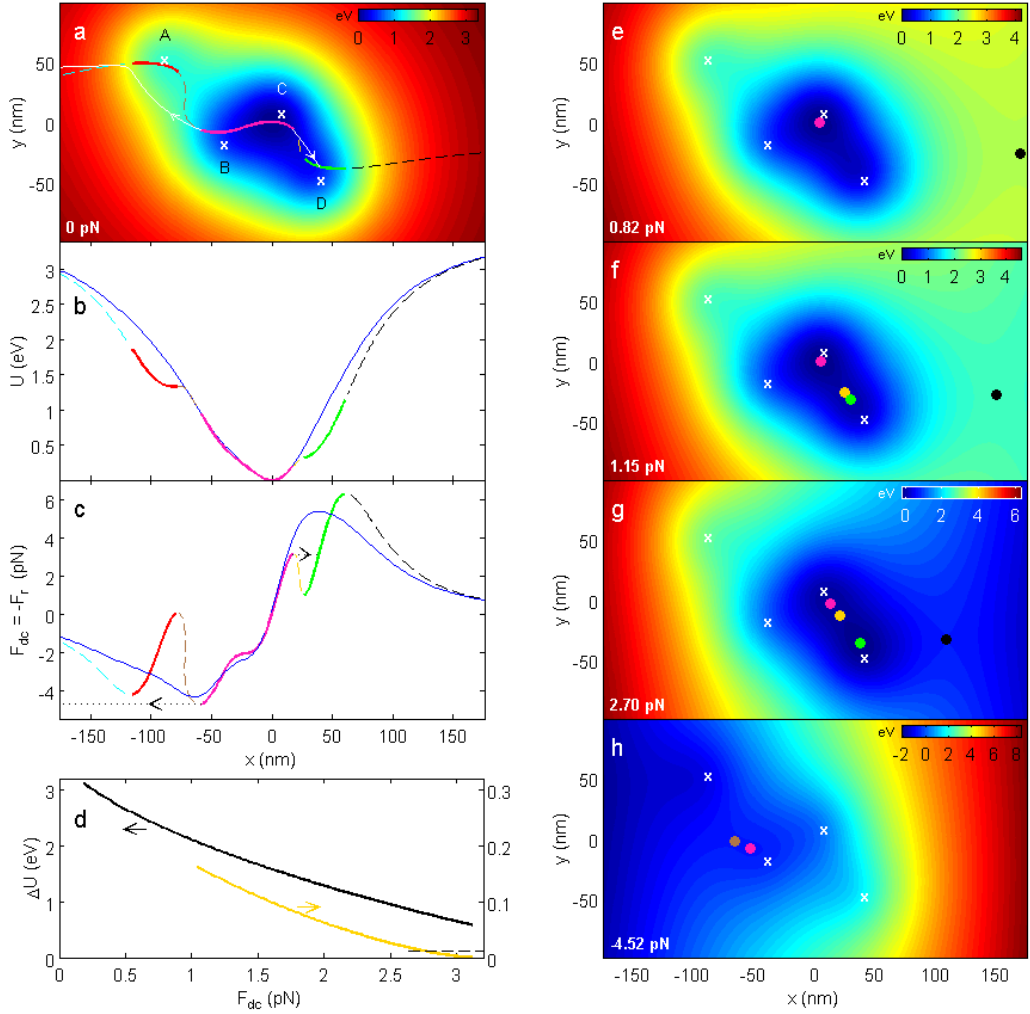

**Figure 5. Simulation of vortex potential and dynamics in a multi-defect well. a**, Calculated 2D vortex potential at zero driving force due to a cluster of four point-defects at locations marked by × (defects A, B, C, and D contribute a Lorentzian potential with $U_0$ of 1.41, 1.41, 2.0, and 2.0 eV respectively). Overlayed is the calculated trajectory of the vortex upon varying the driving force $F_{dc}$: solid color lines – loci of static potential minima points; dashed lines – loci of inflection points; white solid lines – dynamic vortex escape paths out of the central well at positive and negative critical forces. An expanded view of the static trajectory in the central well is shown in Fig. 2j. **b**, $U(x)$ along the stationary vortex trajectory in (**a**) (color solid and dashed segments) as compared to $U(x)$ along the $y = 0$ line in (**a**) (solid blue). **c**, The corresponding restoring force $F_{dc} = -F_r = \partial U/\partial x$. The dashed lines (saddle points in (**a**)) are unstable solutions. The dotted lines show the escape of the vortex at the critical forces out of the central well. **d**, Black: activation barrier between the central minimum (magenta in (**a**)) and the main saddle point (dashed black in (**a**)) vs. $F_{dc}$. Yellow: activation barrier between the central minimum and the metastable saddle point (dashed yellow in (**a**)) that is formed at $F_{dc} > 1.04$ pN. The central minimum disappears for $F_{dc} > 3.1$ pN. The dashed line marks $\Delta U = 34 k_B T = 12$ meV, at which thermal activation becomes relevant. **e**, 2D vortex potential at $F_{dc} = 0.82$ pN at which one minimum (magenta) and one saddle point (black) are present. **f**, 2D potential at $F_{dc} = 1.15$ pN at which two minima (magenta and light green) and two saddle points (yellow and black) are present. **g**, Same as (**f**) at $F_{dc} = 2.70$ pN. For $F_{dc} > 3.1$ pN, only the light green metastable minimum remains. **h**, 2D potential at $F_{dc} = -4.52$ pN, at which only the central minimum (magenta) and a nearby saddle point (brown) are present.

Another ingredient of the spring-breaking is thermal activation, which is usually disregarded for conventional superconductors at low temperatures [11,41]. Figure 5d shows that the typical energy barrier for thermally-activated hopping of vortices between the potential minimum and the main saddle point (black curve) is large ($\Delta U \cong U_0 \cong 1$ eV $\cong 10^4$ K) and decreases smoothly with $F_{dc}$. As a result, thermal activation at 4.2 K becomes relevant only within a few percent of the critical force $F_c$, at which $\Delta U(F_{dc}) \cong 34$ k$_B$T $\cong 12$ meV (see S9 for details). For a single defect, the spring constant at $F_c - F_{dc} \sim 10^{-2} F_c$ is reduced significantly, resulting in $x_{ac}$ that is about five times larger than at the bottom of the well (Figures 2a and S10b) and inconsistent with the experimental data (Figure 2e). However, the metastable wells due to potential ripples create multiple saddle points separated by much smaller activation barriers (yellow line in Fig. 5d) that can cause premature thermal activation of the vortex, facilitating the broken-spring effect (as shown in Figures 2k and 2i). At $J \sim J_c$, the heights of these metastable barriers depend weakly on $F_{dc}$, resulting in thermally-activated nondeterministic hopping of vortices over a wide range of applied currents, consistent with our experimental data. Hopping of vortices over these small potential ripples may also be relevant to the outstanding problem of nearly temperature-independent relaxation rates $s(T,H) = d\ln(J)/d\ln(t)$ of magnetization currents in the critical state of conventional superconductors, for which $s(T)$ remains nearly constant even at $T \ll T_c$ [39,41].

Based on the above insights, we performed a full analysis of a 2D potential comprising a random distribution of defects. Figure 4c shows the potential landscape and the resulting static vortex trajectories inside the wells, as well as the dynamic escape trajectories between the wells (dotted lines). The main features—including the extent of the wells, shape of the trajectories, metastable wells, typical hopping distances and the transverse displacement between the wells—are in good qualitative agreement with the experimental results. In addition, the wells show the broken-spring effect and inflection points in the restoring force (Figure 4d) coinciding with the inflection points in the static trajectories. Consistent with our SOT observations, the vortex trajectories in Figure 4 form closed loops that provide the means for controlled manipulation and braiding of vortices for topological quantum computation [42].

In conclusion, the new scanning SQUID-on-tip microscopy enabled us, for the first time, to measure the fundamental dependence of the elementary pinning forces on vortex displacement. The totality of our experimental and computational results shows that a vortex typically interacts with small clusters of a few pinning defects separated by about the coherence length. At low currents, the random multi-scale pinning landscape results in complex vortex trajectories and unusual softening within the potential wells. On approaching the critical current, the 2D random configuration causes fragmentation of the potential into metastable wells and gives rise to sharp ripples in the restoring force, triggering abrupt thermally-activated depinning of vortices even in the case of strong pinning and low temperatures, for which no significant thermal relaxation is expected. These results provide new insights into the pinning of vortex matter, mechanisms of magnetic relaxation in superconductors at low temperatures, the nonlinear response of superconductors to strong alternating electromagnetic fields, and the development of high-critical-current conductors with artificial pinning nanostructures. This work may also open exciting opportunities in the controllable manipulation of single vortices on nanometer scales, particularly in quantum computations based on braiding and entanglement of vortices in thin film nanostructures.

**Methods**

The SOT that was used for scanning in this work was a Pb-based device [32] with an effective diameter of 177 nm, 103 μA critical current at zero field, and white flux noise down to 230 $n\Phi_0 Hz^{-0.5}$ (S1). The SOT was integrated into a scanning probe microscope with a scanning range of 30×30 μm$^2$ [43] and read out using a series SQUID array amplifier [44]. All measurements were performed at 4.2 K in He exchange gas at a pressure of 0.95 bar.

A 75 nm thick Pb film was deposited by thermal evaporation and capped by 7 nm of Ge. The film was patterned lithographically into an 8 μm wide microbridge fitted with electrical contacts to allow the application of transport currents (S2).

A square wave ac current $I_{ac}$ of 0.56 mA peak-to-peak (ptp) and frequency of 13.3 kHz was applied to the sample, resulting in a Lorentz force $F_{ac} = \Phi_0 J_{ac} = 93.1$ fN ptp on the vortex, where $J_{ac}$ is the corresponding sheet current density at the location of the vortex in the center of the microbridge in the Meissner state [45]. In the softening peak regions, the ac current and the driving force were reduced by a factor of ten, to $F_{ac} = 9.31$ fN, in order to keep $x_{ac}$ below 1 nm. The ac current was superimposed on a dc current $I_{dc}$ in the range of ±27 mA (applying a dc Lorentz force in the range of $F_{dc}$ = ±4.44 pN). The size of step in $I_{dc}$ was kept equivalent to the ptp value of $I_{ac}$ in order to facilitate the displacement integration procedures. At each value of $F_{dc}$, full images of $B_{dc}$ and $B_{ac}$ were acquired simultaneously, with the SOT scanned at a constant height of about 150 nm above the sample. The frame size was typically 2.2×1.3 μm$^2$ with pixel size of 10 nm scanned at a speed of 2 μm/sec, taking about five minutes to complete.

**References**

1. Larbalestier, D., Gurevich, A., Feldmann, D. M. & Polyanskii, A. High-Tc superconducting materials for electric power applications. *Nature* **414**, 368-377 (2001)
2. Macmanus-Driscoll, J. L., Foltyn, S. R., Jia, Q. X., Wang, H., Serquis, A., Civale, L., Maiorov, B., Hawley, M. E., Maley, M. P. & Peterson, D. E. Strongly enhanced current densities in superconducting coated conductors of YBa$_2$Cu$_3$O$_{7-x}$ + BaZrO$_3$, *Nature Mater.* **3**, 439-443 (2004)
3. Fang L., Jia, Y., Mishra, V., Chaparro, C., Vlasko-Vlasov, V. K., Koshelev, A. E., Welp, U., Crabtree, G. W., Zhu, S., Zhigadlo, N. D., Katrych, S., Karpinski, J. & Kwok, W. K. Huge critical current density and tailored superconducting anisotropy in SmFeAsO$_{0.8}$F$_{0.15}$ by low-density columnar-defect incorporation. *Nat Commun.* **4**, 2655 (2013)
4. Haugan, T. J., Barnes, P. N., Wheeler, R., Meisenkothen, F. & Sumption, M. D. Addition of nanoparticle dispersions to enhance flux pinning in the YBa$_2$Cu$_3$O$_{7-x}$ superconductor. *Nature* **430**, 867-870 (2004)
5. Mele, P., Matsumoto, K., Horide, T., Miura, O., Ichnose, A., Mukaida, M., Yoshida, Y. & Horii, S. Tuning of the critical current in YBa$_2$Cu$_3$O$_{7-x}$ thin films by controlling the size and density of Y$_2$O$_3$ nanoislands on annealed SrTiO$_3$ substrates. *Supercond. Sci. Technol*. **19**, 44-50 (2006)

6. Kang, S., Goyal, A., Li, J., Gapud, A. A., Martin, P. M., Heatherly, L., Thompson, J. R., Christen, D. K., List, F. A., Paranthaman, M. & Lee, D. F. High-performance high-$T_c$ superconducting wires. *Science* **311**, 1911-1914 (2006)
7. Gutierrez, J., Lloreds, A., Gazquez, J., Gibert, M., Roma, N., Ricart, S., Pomar, A., Sandiumenge, F., Mesters, N., Puig, T. & Obradors, X. Strong isotropic flux pinning in solution-derived $YBa_2Cu_3O_{7-x}$ nanocomposite superconducting films. *Nature Mater.* **6**, 367-373 (2007)
8. Maiorov, B., Baily, S. A., Zhou, H., Ugurlu, O., Kennison, J. A., Dowden, P. C., Holesinger, T. G., Foltyn, S. R. & Civale, L. Synergetic combination of different types of defects to optimize pinning landscape using $BaZrO_3$ - doped $YBa_2Cu_3O_{7-x}$. *Nature Mater.* **8**, 398-404 (2009)
9. Llordés, A., Palau, A., Gázquez, J., Coll, M., Vlad, R., Pomar, A., Arbiol, J., Guzmán, R., Ye, S., Rouco, V., Sandiumenge, F., Ricart, S., Puig, T., Varela, M., Chateigner, D., Vanacken, J., Gutiérrez, J., Moshchalkov, V., Deutscher, G., Magen, C. & Obradors X. Nanoscale strain-induced pair suppression as a vortex-pinning mechanism in high-temperature superconductors. *Nature Mater.* **11**, 329-36 (2012)
10. Lee, S., Tarantini, C., Gao, P., Jiang, J., Weiss, J. D., Kametani, F., Folkman, C. M., Zhang, Y., Pan, X. Q., Hellstrom, E. E., Larbalestier, D. C. & Eom C. B. Artificially engineered superlattices of pnictide superconductors. *Nature Mater.* **12**, 392-396 (2013)
11. Blatter, G., Feigel'man M. V., Geshkenbein V. B., Larkin A. I. & Vinokur V. M. Vortices in high-temperature superconductors. *Rev. Mod. Phys.* **66,** 1125 (1994)
12. Lee, C. S., Jankó, B., Derényi, I. & Barabási, A. L. Reducing vortex density in superconductors using the 'ratchet effect'. *Nature* **400**, 337-340 (1999)
13. Villegas, J. E., Savel'ev, S., Nori, F., Gonzalez, E. M., Anguita, J. V., García, R. & Vicent J. L. Superconducting reversible rectifier that controls the motion of magnetic flux quanta. *Science* **302**, 1188-1191 (2003)
14. Zhu, B. Y., Marchesoni, F. & Nori, F. Controlling the Motion of Magnetic Flux Quanta. *Phys. Rev. Lett.* **92**, 180602 (2004)
15. de Souza Silva, C. C., Van de Vondel, J., Morelle, M. & Moshchalkov, V. V. Controlled multiple reversals of a ratchet effect. *Nature* **440**, 651-654 (2006)
16. Lu, Q., Olson Reichhardt, C. J. & Reichhardt. C. Reversible vortex ratchet effects and ordering in superconductors with simple asymmetric potential arrays. *Phys. Rev. B* **75**, 054502 (2007)
17. Oral A., Barnard, J. C., Bending, S. J., Kaya, I. I., Ooi, S., Tamegai, T. & Henini, M. Direct observation of melting of the vortex solid in $Bi_2Sr_2CaCu_2O_{8+\delta}$ single crystals, *Phys. Rev. Lett.* **80**, 3610 (1998)
18. Troyanovski, A. M., Aarts, J. & Kes, P. H. Collective and plastic vortex motion in superconductors at high flux densities. *Nature* **399**, 665-668 (1999)
19. Bending, S. J. Local magnetic probes of superconductors, *Advances in Physics* **48**, 449-535 (1999)
20. Kirtley, J. R. Fundamental studies of superconductors using scanning magnetic imaging, *Rep. Prog. Phys.* **73**, 126501 (2010)
21. Auslaender, O. M., Luan, L., Straver, E. W. J., Hoffman, J. E., Koshnick, N. C., Zeldov, E., Bonn, D. A., Liang, R., Hardy, W. N. & Moler, K. A. Mechanics of individual isolated vortices in a cuprate superconductor. *Nature Phys.* **5**, 35-39 (2009)

22. Kalisky, B., Kirtley, J. R., Nowadnick, E. A., Dinner, R. B., Zeldov, E., Ariando, Wenderich, S., Hilgenkamp, H., Feldmann, D. M. & Moler, K. A. Dynamics of single vortices in grain boundaries: I-V characteristics on the femtovolt scale. *Appl. Phys. Lett.* **94**, 202504 (2009)
23. Guillamon, I., Suderow, H., Fernandez-Pacheco, A., Sese, J., Cordoba, R., De Teresa, J. M., Ibarra, M. R. & Vieira, S. Direct observation of melting in a two-dimensional superconducting vortex lattice. *Nature Physics* **5**, 651 - 655 (2009)
24. Lee, J., Wang, H., Dreyer, M., Berger, H. & Barker, B. I. Nonuniform and coherent motion of superconducting vortices in the picometer-per-second regime. *Phys. Rev. B* **84**, 060515 (2011)
25. Kalisky, B., Kirtley, J. R., Analytis, J. G., Chu, J. H., Fisher, I. R. & Moler, K. A. Behavior of vortices near twin boundaries in underdoped Ba(Fe$_{1-x}$Co$_x$)$_2$As$_2$. *Phys. Rev. B* **83**, 064511 (2011)
26. Raes, B., Van de Vondel, J., Silhanek, A. V., de Souza Silva, C. C., Gutierrez, J., Kramer, R. B. G. & Moshchalkov, V. V. Local mapping of dissipative vortex motion. *Phys. Rev. B* **86**, 064522 (2012)
27. Timmermans, M., Samuely, T., Raes, B., Van de Vondel, J. & Moshchalkov, V. V. Dynamic Visualization of Nanoscale Vortex Orbits. *ACS Nano* **8**, 2782–2787 (2014)
28. Suderow, H., Guillamón, H., Rodrigo, J. G. & Vieira, S. Imaging superconducting vortex cores and lattices with a scanning tunneling microscope. *Supercond. Sci. Technol.* **27**, 063001 (2014)
29. Ganguli, S. G., Saraswat, G., Ganguly, R., Singh, H., Bagwe, V., Shirage, P., Thamizhavel, A. & Raychaudhuri, P. Direct evidence of two-step disordering of the vortex lattice in a 3 dimensional superconductor, Co0.0075NbSe2. arXiv:1406.7422
30. Raes, B., de Souza Silva, C. C., Silhanek, A. V., Cabral, L. R. E., Moshchalkov, V. V. & Van de Vondel, J. A closer look at the low frequency dynamics of vortex matter. arXiv:1406.3939
31. Finkler, A., Segev, Y., Myasoedov, Y., Rappaport, M. L., Ne'eman, L., Vasyukov, D., Zeldov, E., Huber, M. E., Martin, J. & Yacoby, A. Self-aligned nanoscale SQUID on a tip. *Nano Lett.* **10**, 1046-1049 (2010)
32. Vasyukov, D., Anahory, Y., Embon, L., Halbertal, D., Cuppens, J., Neeman, L., Finkler, A., Segev, Y., Myasoedov, Y., Rappaport, M. L., Huber, M. E. & Zeldov, E. A scanning superconducting quantum interference device with single electron spin sensitivity. *Nature Nanotech.* **8**, 639-644 (2013)
33. Bespalov, A.A. & Mel'nikov, A.S. Abrikosov vortex pinning on a cylindrical cavity inside the vortex core: formation of a bound state and depinning. *Supercond. Sci. Technol.* **26,** 085014 (2013)
34. Gurevich, A. & Cooley, L.D. Anisotropic flux pinning in a network of planar defects. *Phys. Rev.* B **50**, 13563 (1994)
35. Thuneberg, E. V., Kurkijarvi, J. & Rainer, D. Elementary-flux-pinning potential in type-II superconductors. *Phys. Rev.* B **29**, 3913 (1984)
36. Blatter, G., Geshkenbein, V. B. & Koopmann, J. A. G. Weak to strong pinning crossover. *Phys. Rev. Lett.* **92**, 067009 (2004)
37. Koshelev, A. E. & Kolton, A. B. Theory and simulations on strong pinning of vortex lines by nanoparticles. *Phys. Rev.* B **84**, 104528 (2011)
38. Thomann, A. U., Geshkenbein, V. B., & Blatter, G. Dynamical aspects of strong pinning of magnetic vortices in type-II superconductors, *Phys. Rev. Lett.* **108**, 217001 (2012)
39. Campbell, A. M. & Evetts, J. E. Flux vortices and transport current in type-II superconductors. *Adv. Phys.* **21**, 194–428 (1972)

40. Priour, D. J. & Fertig, H. A. Deformation and depinning of superconducting vortices from artificial defects: A Ginzburg-Landau study. *Phys. Rev.* B **67**, 054504 (2003)
41. Yeshurun, Y., Malozemoff, A. P. & Shaulov, A. Magnetic relaxation in high-temperature superconductors. *Rev. Mod. Phys*. **68**, 911-949 (1996)
42. Nayak, C., Simon, S. H. Stern, A., Freedman, M. & Das Sarma, S. Non-Abelian anyons and topological quantum computation. *Rev. Mod. Phys.* **80**, 1083 – 1159 (2008)
43. Finkler, A., Vasyukov, D., Segev, Y., Ne'eman, L., Lachman, E. O., Rappaport, M L., Myasoedov, Y., Zeldov, E. & Huber M. E. Scanning superconducting quantum interference device on a tip for magnetic imaging of nanoscale phenomena. *Rev. Sci. Instrum.* **83**, 073702 (2012)
44. Huber, M.E., Neil, P.A., Benson, R.G., Burns, D.A., Corey, A. F., Flynn, C. S., Kitaygorodskaya, Y., Massihzadeh, O., Martinis, J. M. & Hilton, G. C. DC SQUID series array amplifiers with 120 MHz bandwidth. *IEEE Trans. Appl. Supercond.* **11**, 4048 (2001)
45. Zeldov, E., Clem, J. R., McElfresh, M. & Darwin, M. Magnetization and transport currents in thin superconducting films. *Phys. Rev. B* **49**, 9802-9822 (1994)

**Acknowledgments**

This work was supported by the US-Israel Binational Science Foundation (BSF), the European Research Council (ERC advanced grant), and by the Minerva Foundation with funding from the Federal German Ministry for Education and Research. E.Z. acknowledges support by the Israel Science Foundation (grant No. 132/14). Y.A. acknowledges support by the Azrieli Foundation and by the Fonds Québécois de la Recherche sur la Nature et les Technologies. M.H acknowledges support from a Fulbright Fellowship awarded by the United States-Israel Educational Foundation. This study was made possible by the able hands of the late S. Sharon who hand-crafted the parts for the scanning microscope.

**Author contribution**

L.E., Y.A. and E.Z. developed and carried out the experiment. Y.A., M.R. and Y.M. developed the SOT fabrication technique. Y.A. and A.U. fabricated the SOT used in this work. L.E. designed and constructed the scanning SOT microscope.  M.E.H. developed the SOT measurement system. L.E. and Y.A. designed and fabricated the sample. Y.A., J.C. and L.E. characterized the sample. Y.A. analyzed the data. A.Y. contributed to the analysis software. J.C. contributed to the experiment and data interpretation. Y.A., A.S. and A.G. carried out and analyzed numerical and analytical calculations. D.H. performed numerical analysis and parametric fitting. E.Z., Y.A., L.E. and A.G. co-wrote the manuscript. All authors contributed to the manuscript.

**Competing Financial Interests statement**

The authors declare no competing financial interests.

# Supplementary material

# Probing dynamics and pinning of single vortices in superconductors at nanometer scales

L. Embon[1,*], Y. Anahory[1,*], A. Suhov[1], D. Halbertal[1], J. Cuppens[1], A. Yakovenko[1], A. Uri[1], Y. Myasoedov[1], M.L. Rappaport[1], M.E. Huber[2], A. Gurevich[3] and E. Zeldov[1]

[1]*Department of Condensed Matter Physics, Weizmann Institute of Science, Rehovot, 7610001, Israel*
[2]*Department of Physics, University of Colorado Denver, Denver, 80217, USA*
[3]*Department of Physics, Old Dominion University, Norfolk, VA 23529-0116, USA*
[*]*These authors contributed equally to this work*
*Corresponding authors: lior.embon@weizmann.ac.il, yonathan.anahory@weizmann.ac.il*

1. **SQUID-on-tip (SOT) characteristics**

A detailed explanation and description of the scanning SQUID-on-tip microscopy method, including the fabrication process of the SOTs, their characterization, and the measurement technique, can be found in Refs. 31,32, and 43. The SOT used for this work was a Pb-based device with an effective diameter of 177 nm (83.7 mT field modulation period), 103 µA critical current at zero field, and white flux noise (at frequencies above a few hundred Hz) down to $230\ n\Phi_0 Hz^{-0.5}$. All the measurements were performed at 4.2 K in He exchange gas at 0.95 bar. Figure S1 shows the measured quantum interference pattern for this device. The asymmetric structure of this SOT provides high sensitivity even at zero applied field with flux noise of $360\ n\Phi_0 Hz^{-0.5}$.

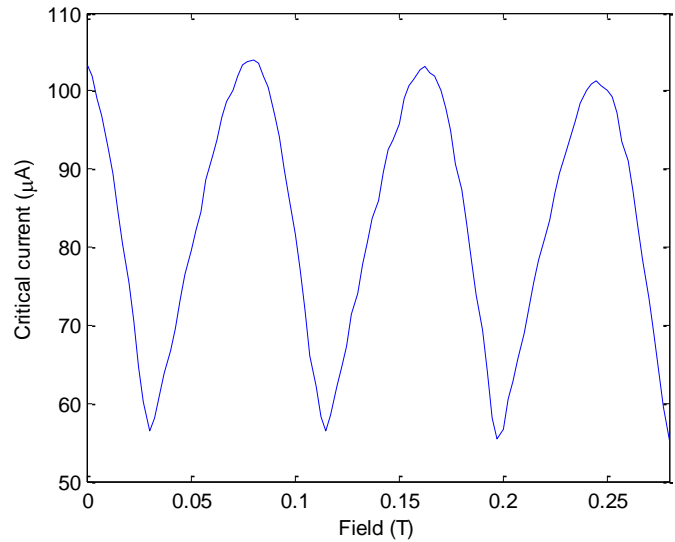

**Figure S1. $I_c$ of the SOT that was used in this work as a function of applied magnetic field.** The asymmetric structure of the SOT results in sensitivity at zero applied field.

2. **Sample fabrication and characteristics**

A 75 nm thick Pb film was deposited by thermal evaporation on a silicon substrate cooled to liquid nitrogen temperature in order to prevent island growth that would otherwise result from the high mobility of the Pb atom. The base pressure was 2.2×10$^{-7}$ Torr and the deposition rate was 0.6 nm/s. A protective layer of 7 nm of Ge was deposited *in-situ* to prevent oxidation of the Pb. The sample was patterned using a standard lift-off lithographic process. To characterize the film, several techniques were employed.

2a. **Scanning Electron Microscopy (SEM)**

Figure S2a shows a SEM image of the sample. Six micro-bridges of different widths (ranging from 5 to 20 µm) were patterned in the film, through which individual currents could be applied. The measurements presented here were performed on the third bridge from the right, which is shown in more detail in Figure S2b. The central straight part of the bridge is 8 µm wide and 12 µm long. Figure S2c shows a zoomed-in image of the film exhibiting a granular structure typical of metallic surfaces. The grain diameter is on the order of a few tens of nm. No other distinct defects were observed

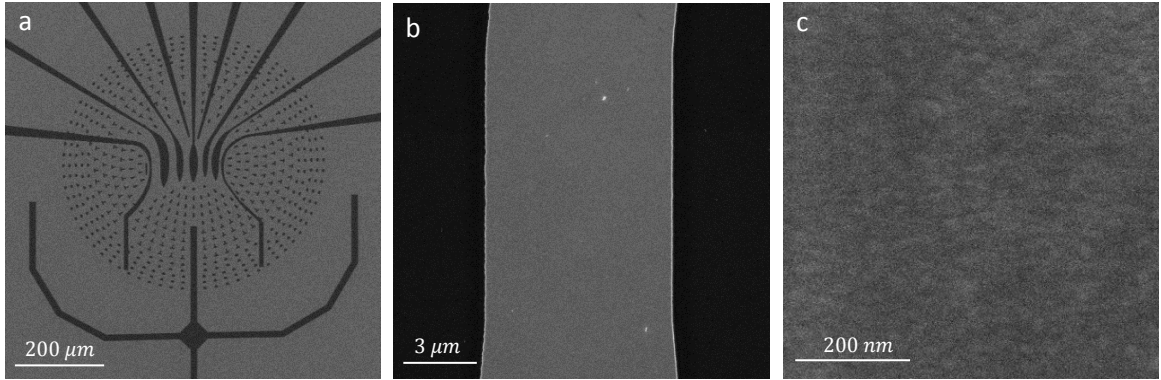

**Figure S2. SEM images of the sample.** The Pb film is light gray and the substrate is dark. **a,** Six micro-bridges with different widths ($5 - 20\ \mu m$) are patterned in the center. Currents can be applied through each of the bridges independently. A circular pattern of shaped holes provides a map for SOT microscope navigation. **b,** A zoomed-in image of the bridge on which the measurements were performed. The central straight section is 8 µm wide and 12 µm long. **c,** Surface of the Ge-capped Pb film showing grains with a typical diameter of a few tens of nm.

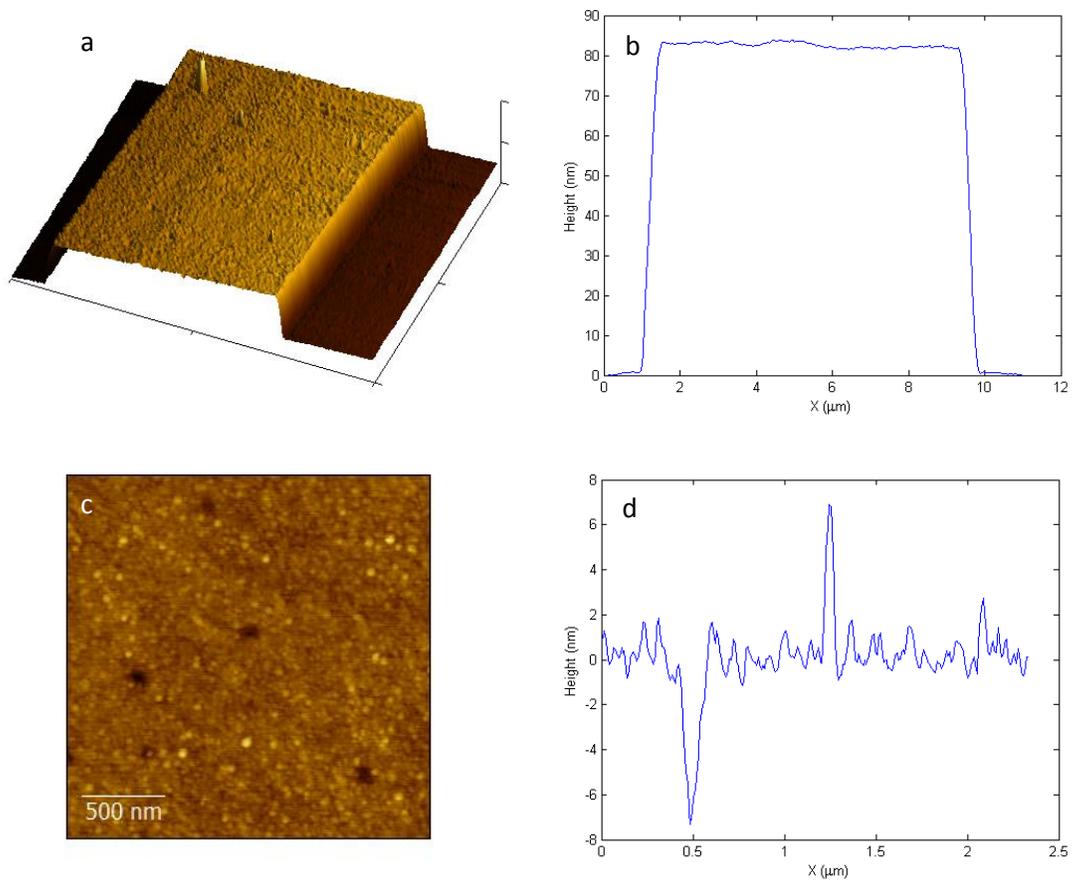

**Figure S3. AFM scans of the 8 $\mu m$ wide bridge. a,** A 3D representation of a 13x13 $\mu m^2$ scan. The surface of the film has granular features, as expected for a metallic surface. The patterned edges of the bridge are clear and sharp. **b,** An averaged cross section of the scan. The overall thickness of the film is 82 nm, which consists of 75 nm of Pb capped by 7 nm of Ge. **c,** A 2x2 $\mu m^2$ scan of the surface of the 8 um wide bridge. The false color scale spans 16 nm. The surface of the film is granular with a typical grain diameter on the order of a few tens of nm. **d,** An extracted line profile that was chosen to include both the highest and the lowest measured regions of **(c)**. The calculated root mean square roughness is 1 nm.

## 2b. Atomic Force Microscopy (AFM)

Figure S3a shows a 3D representation of a 13x13 $\mu m^2$ scan of the 8 $\mu m$ wide bridge using a commercial AFM. The overall thickness of the film was measured to be 82 nm which consists of a 75 nm thick Pb film and a 7 nm thick Ge capping layer. An averaged cross section of that scan is shown in Figure S3b.

Figure S3c is a high resolution 2x2 $\mu m^2$ scan of the surface showing a typical grain diameter on the order of a few tens of nm. Figure S3d displays a line cut through this scan showing a root mean square roughness of the surface of 1 nm.

## 2c. X-ray Diffraction (XRD) measurements

The crystallographic phases of the Pb film were identified using XRD. An asymmetric 2θ scan with a fixed incident angle at 2 degrees was performed. The sample consists of a 70 nm thick unpatterned Pb film capped with 7 nm of Ge to prevent oxidation and was grown in similar conditions to the sample used in our experiment. Figure S4 shows ten Pb peaks associated with different (hkl) orientations of the crystallites. The unmarked peak corresponds to the Si substrate (asymmetrical (311)). The 7 nm Ge layer was not detected. It should be noted that the small incident angle results in a higher diffraction intensity from the thin film than from the substrate. Moreover, the direction of the observed lattice plain varies with the 2θ position, revealing the polycrystalline nature of the studied films, without any preferred orientation of crystallites. Using the Scherrer formula, the grain size was estimated to be 57 nm.

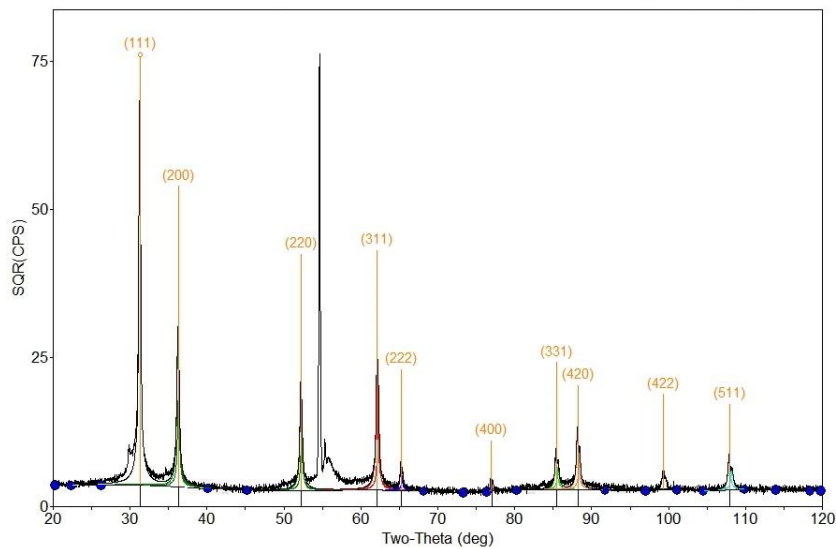

**Figure S4. X-ray diffraction pattern of a Pb film.** The measurement was performed on a 70 nm thick Pb film grown under similar conditions as the sample. Ten different crystallite orientations were identified in addition to the signal from the Si substrate (unmarked peak).

## 2d. Electrical transport measurement

Transport measurements exploring the upper critical field $H_{c2}(T)$ were conducted in order to determine the coherence length $\xi$ and the penetration depth $\lambda$ of the superconducting film. Figure

S5a shows the resistance of the bridge as a function of the applied magnetic field for various temperatures.

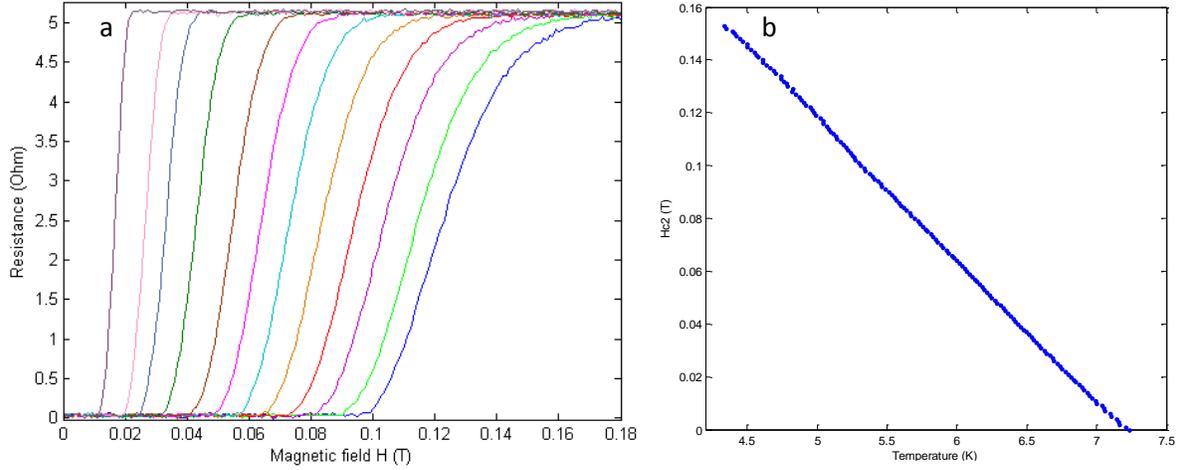

**Figure S5. Transport measurement of the sample. a,** Resistance vs. applied magnetic field at different temperatures. The temperature values are equally spaced between 6.8 K (purple) and 4.44 K (blue). **b,** Superconducting upper critical field $\boldsymbol{H_{c2}(T)}$. The line was determined using the $\frac{\boldsymbol{R}}{\boldsymbol{R_N}} = \boldsymbol{0.9}$ criterion, where $\boldsymbol{R_N}$ is the normal state resistance.

Figure S5b shows the extracted $H_{c2}(T)$ line, determined at 90% of the normal state resistance. From the measured $H_{c2}(4.2 \text{ K}) = 153 \; mT$, and using the expression $H_{c2}(T) = \frac{\Phi_0}{2\pi\xi(T)^2}$ we find the coherence length $\xi(4.2 \text{ K}) = 46 \; nm$. Assuming that the thermodynamic critical field $H_c$ is unaffected by disorder, the magnetic penetration depth $\lambda(4.2K)$ can be calculated using the expression $H_c(T) = \frac{\Phi_0}{2 \sqrt{2} \pi \lambda(T)\xi(T)}$. With $H_c(4.2K) = 53 \; mT$ for Pb we find $\lambda(4.2K) = 96 \; nm$.

### 3. Single-vortex state preparation

To prepare the bridge in a single-vortex state, a quench and field-cooling technique was used. A current of a few tens of mA was applied to the microbridge which drove it to the normal state and heated it above $T_c$. The current was then turned off abruptly, and the film was field-cooled in an applied field of about 1 G. Due to a small remnant field of the superconducting magnet, the exact nominal value of the applied field required to trap a single vortex was found by trial and error. Figure S6 shows four 12×10 $\mu m^2$ scans of the same 8 $\mu m$ wide bridge with 0, 2, 6 and 15 trapped vortices after field cooling in 0.7, 1, 3 and 7 G, respectively.

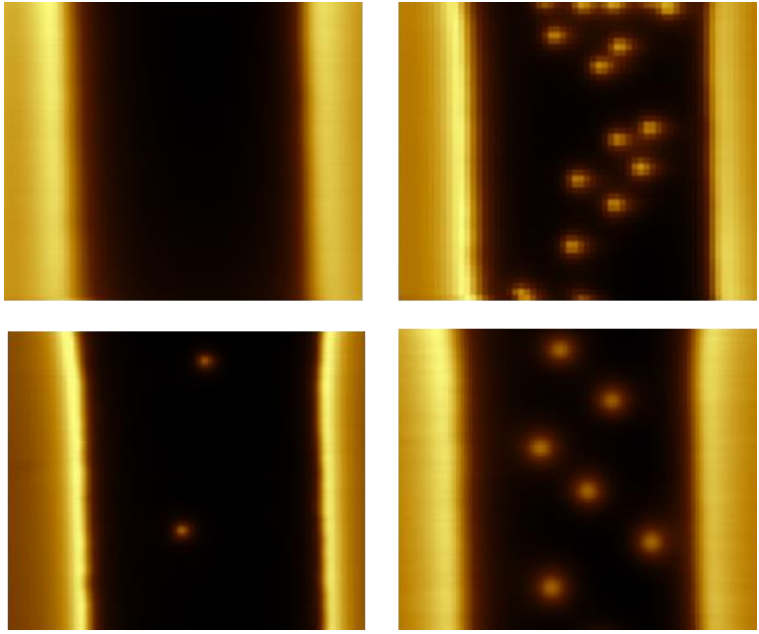

**Figure S6. $B_{dc}$ scans after field cooling.** The 12×10 $\mu m^2$ images show 0, 2, 6 and 15 vortices in the same area of the $8\ \mu m$ wide bridge after field cooling in fields of 0.7, 1, 3 and 7 G, respectively.

## 4. Sensitivity of vortex displacement measurements

The sensitivity of the measurement of $x_{ac}$ and $y_{ac}$ using the procedure described in Figure 1 is determined mainly by the very high sensitivity of the SOT rather than by its physical size (provided it is sufficiently small). As a result, although an SOT with a diameter of 177 nm was used in the measurement, displacements with sub-angstrom precision could be obtained. If the vortex ac displacement is mainly along the $x$ direction, there is no need to take a full 2D image, and a single line scan measurement through the center of the vortex is sufficient to determine the displacement $x_{ac}$. Figure S7a shows such a measurement of $x_{ac}$ as a function of the excitation force $F_{ac}$, resolving displacements as low as 10 pm.

Note that, since the vortex displacement inside the potential well is essentially dissipationless the ac response showed no frequency dependence up to our maximum lock-in amplifier (LIA) frequency of 100 kHz and had no out-of-phase component. The ac response is linear for vortex displacements of up to a few nm (Fig. S7a). The response becomes nonlinear at larger amplitudes, giving rise to higher harmonics. All the presented data were acquired in the linear regime.

Since we measure both $B_{dc}$ and $B_{ac}$ images at each $F_{dc}$ value, the vortex displacement and the trajectory could, in principle, also be derived from the $B_{dc}$ image by tracing the position of the maximum field using data interpolation. This method, however, is much less sensitive since it depends on the pixel size of the image and on the size of the SOT, and since it is strongly influenced by thermal drifts in the microscope. The ac method, in contrast, is immune to the drifts and can resolve displacements that are some three orders of magnitude smaller than the pixel size. Figure S7b shows a comparison between the two methods demonstrating the advantage of the ac method. Note that the pixel size was 10×10 nm² so the width of the entire plot is just about 5 pixels.

Finally, to demonstrate the reproducibility of the measurement, Figures S7c and S7d show six independent runs of measurements of the ac displacements $x_{ac}$ and $y_{ac}$ in well 5. The data clearly shows the very high reproducibility and consistency of the measurements and of the derivation method.

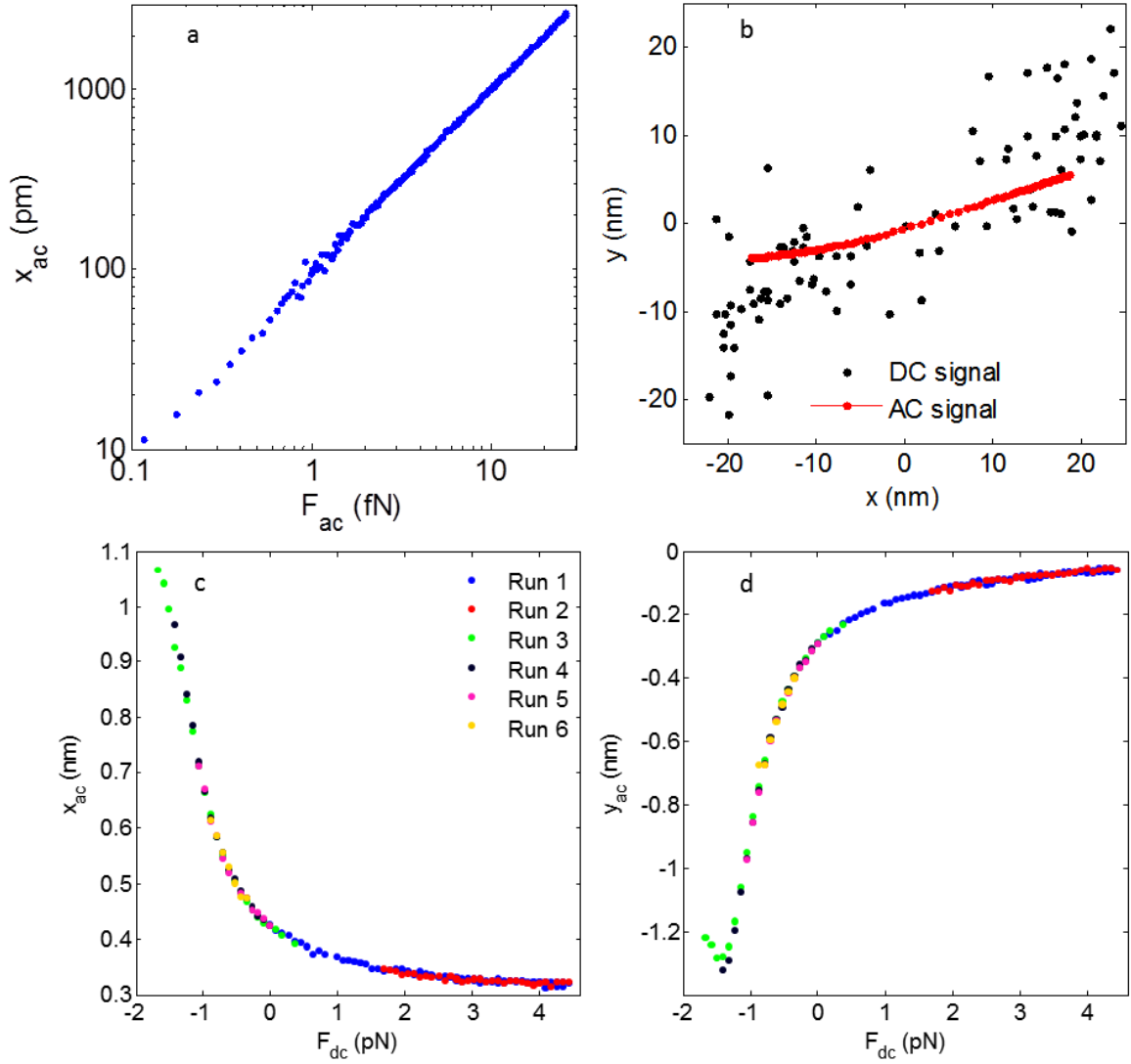

**Figure S7. Sensitivity of vortex displacement measurements. a,** Measured vortex displacement $x_{ac}$ as a function of the ac driving force $F_{ac}$. The measurement demonstrates the ability to resolve displacements as small as 10 pm. The time constant of the LIA was 10 ms for $F_{ac}$ > 1 fN and up to 300 ms at lower drive amplitudes. **b,** Vortex trajectory within well 4 upon varying $F_{dc}$. The two data sets were extracted from the same experimental run using two methods: the DC signal was derived from the peak position of the $B_{dc}$ image at each $F_{dc}$, and the AC signal was derived by integration of $x_{ac}$ as described in the main text. **c-d,** $x_{ac}$ and $y_{ac}$ as a function of $F_{dc}$ within well 5. The data were acquired in six different runs at different times yet the extracted displacements show remarkable consistency. The ac driving force was $F_{ac} = 88.9$ fN.

## 5. Current distribution in the microbridge

We need to know the sheet current density $J$ at the position of the vortex in order to calculate the Lorentz force on the vortex $F = \Phi_0 J$. Since the microbridge is in the Meissner state, the distribution of the applied transport current across the width $w = 8$ µm of the bridge is given by $J(x) = I/\left(\pi\sqrt{(w/2)^2 - x^2}\right)$, where $I$ is the total applied current [45]. Because our vortex is essentially in the center of the bridge ($x = 0$), the sheet current density is $J(0) = 2I/\pi w$ and varies by less than 1% over the range of $\pm 0.5$ μm that the vortex moves during the experiment. For $I = 1$ mA the sheet current density in the center of the bridge is $J(0) = 79.6$ A/m and it exerts a driving force of $F = 165$ fN on the vortex.

We also applied a field of $B_a = 0.3$ mT perpendicular to the bridge. This field induces a Meissner sheet current density of $J(x) = -B_a x / \left(2\mu_0 \pi \sqrt{(w/2)^2 - x^2}\right)$ [45] which is zero in the center of the bridge and reaches $J = 4.8$ A/m and $F = 9.9$ fN at the maximum vortex displacement of about 0.5 μm from the center. This dc force pushing the vortex towards the center of the bridge is negligible compared to our applied dc force of 4.44 pN. It can be readily incorporated into the calculations but was neglected for simplicity.

## 6. Movie of vortex dynamics

The movie comprises about 200 $B_{ac}$ images, corresponding to a closed loop sweeping of $F_{dc}$ from -4.44 pN to 4.44 pN and back at a constant $F_{ac} = 93.1$ fN. The intensity of the dipole-like signal reflects the amplitude of the ac vortex displacement while the orientation of the dipole shows the direction of the vortex motion.

On the ascending branch of the $F_{dc}$ loop, the vortex is initially located in well 1 and then hops to wells 2, 6, and 5, where it stays up to the maximum value of $F_{dc}$. On the descending $F_{dc}$ branch, it is seen that well 5 is initially very stiff, resulting in very small ac displacements. As $F_{dc}$ decreases the well becomes significantly softer and the orientation of the ac displacement changes substantially. The vortex then hops to well 4 and well 3, and eventually to well 2 (without returning to well 1), closing a counter-clockwise hysteretic trajectory in space. Note that well 2 is quite extended and that the vortex explores different and non-overlapping regions of the well on ascending and descending $F_{dc}$ sweeps. In order to explore the full trajectory within well 2, a subloop sweep of $F_{dc}$ was performed. Since some of the vortex jumps are non-deterministic, every repetition of the loop may result in a slightly different sequence of jumps between wells 1 to 6 as described in Figure 4.

## 7. Additional potential wells and closed loop trajectories

Numerous sets of pinning wells were measured as part of this work. Here we present an additional example of a closed-loop trajectory comprising four potential wells. Figure S8a and S8b show a full map of the closed loop vortex trajectories and their corresponding restoring forces. Figures S8c-e show a comparison between the trajectories within each well, the restoring forces, and the potential structures.

Notice that the features discussed in the main text also appear in this set, including the non-trivial trajectories and anisotropy of the wells, inflection points of the restoring force away from the edges, the broken spring effect, the non-deterministic hopping behavior, and the appearance of metastable wells.

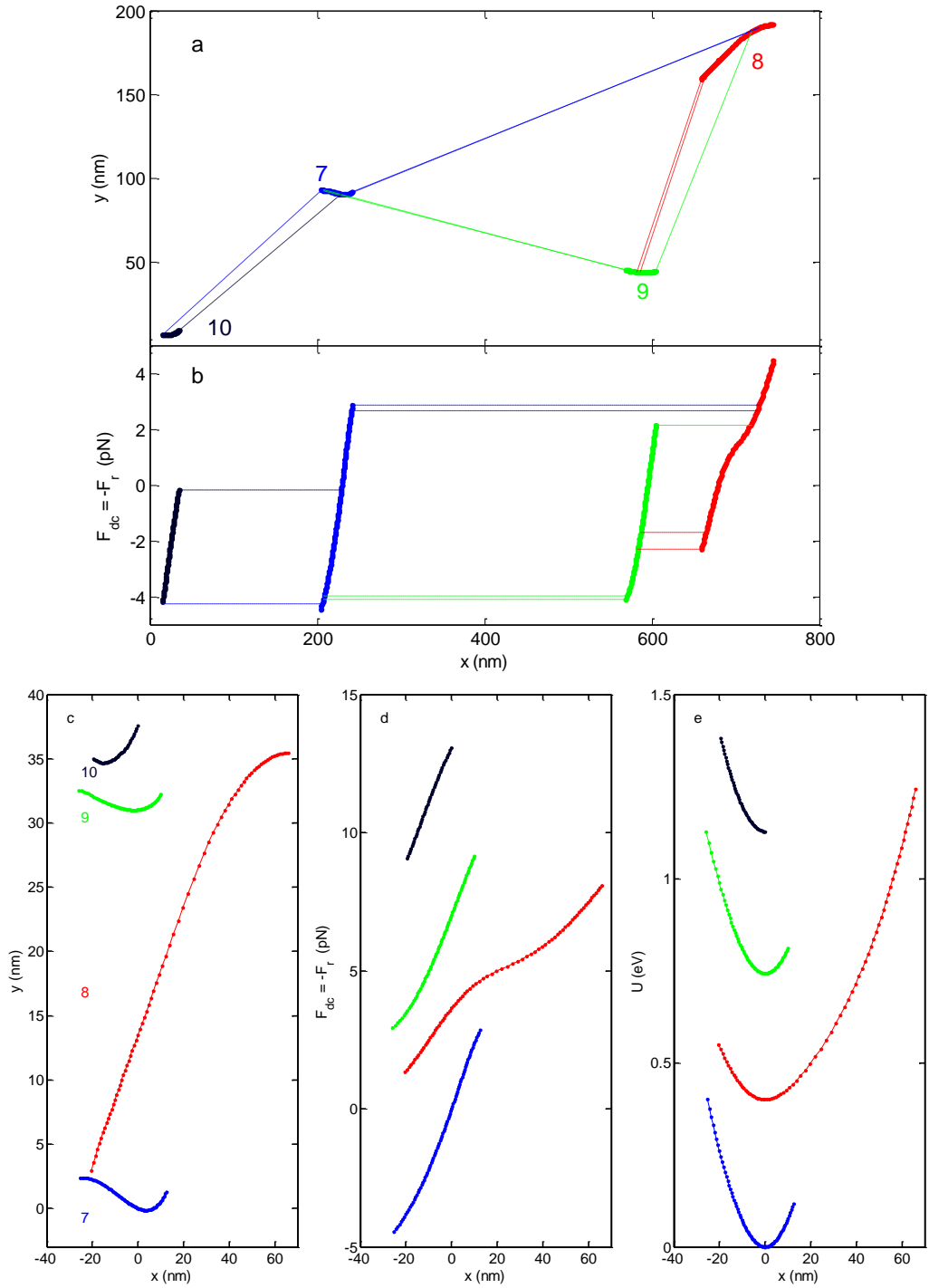

**Figure S8. Additional set of potential wells a**, Closed loop trajectory with vortex trajectories in wells 7 to 10 (solid symbols) and hopping events between the wells (shown schematically as dashed lines with a color matching the original well). The hopping between the wells results in hysteretic closed-loop trajectories that vary upon repeated different cycles of the full loop and of the sub-loops. **b**, The restoring force of the wells $F_{dc} = -F_r(x)$ with hopping events shown schematically (dashed lines with a color matching the original well). The vortex jumps to the right (left) when a positive (negative) applied force exceeds the restoring force. The vortex stops at a position where the restoring force of the new well equals the applied force (horizontal dashed lines). Well 7 is metastable and exists only at negative forces. **c-e**, Comparison of additional potential wells **c**, Vortex trajectories in wells 7 to 10, shifted vertically for clarity. $x = 0$ corresponds to the resting position of the vortex at $F_{dc} = 0$, except for well 7 which does not exist at $F_{dc} = 0$. All the wells display nontrivial

internal structure. **d**, The restoring force $F_{dc} = -F_r$ of the different wells shifted vertically for clarity. $x = 0$ corresponds to $F_{dc} = 0$ for each well except well 4. **e**, The structure of the potential of the different wells, shifted for clarity.

## 8. Softening in the middle of the well due to two defects

The experimentally observed large softening in the ac vortex response in the middle of the well can be explained by two (or more) nearby point defects that form a single potential well as described in Figure S9. For two identical defects the softening occurs exactly in the center of the well at the minimum of the potential. If the defects have different $U_0$ or if there are additional nearby defects, the softening point will be shifted away from the center.

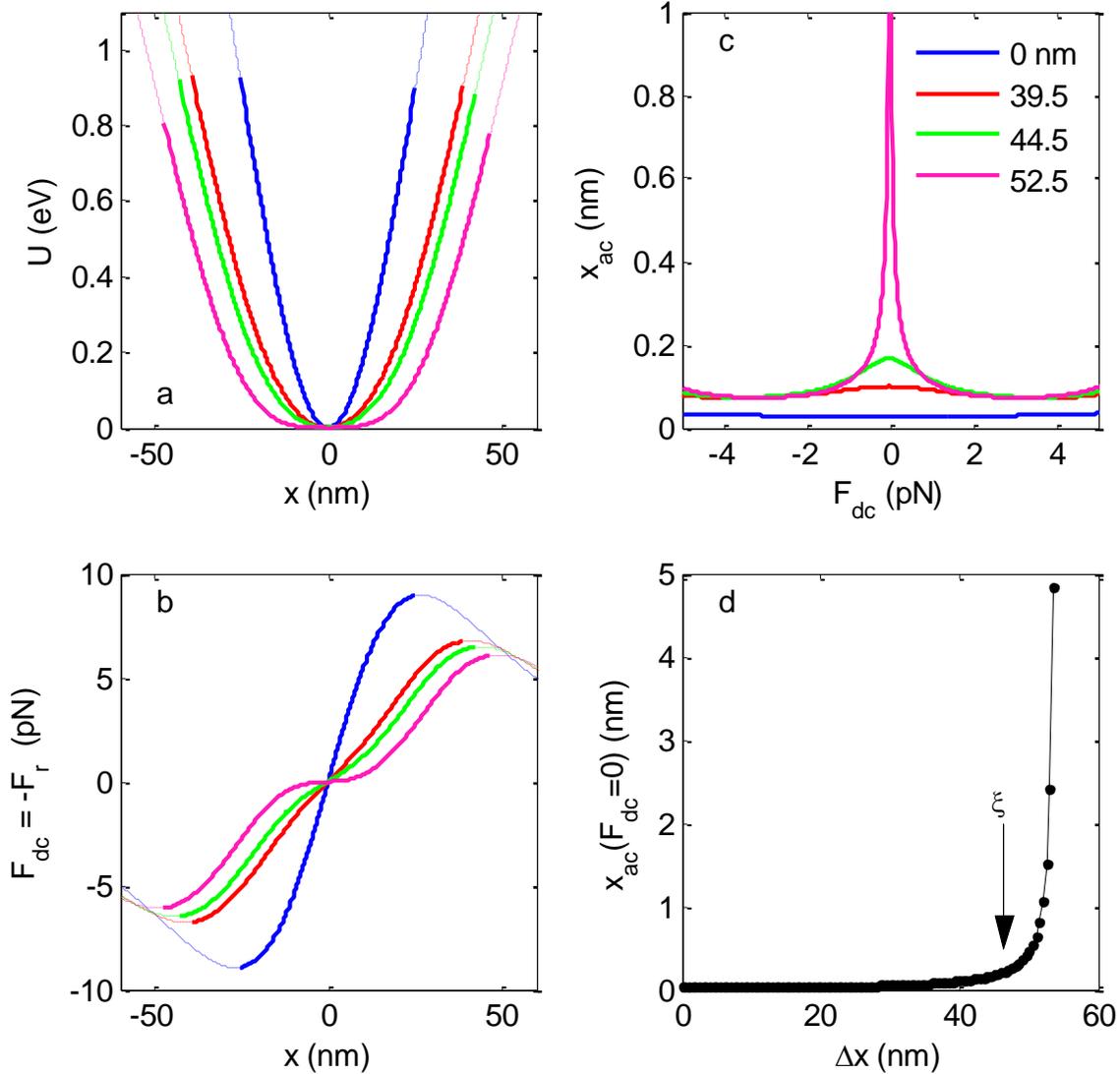

**Figure S9. Potential structure and vortex response in the case of two defects separated by $\Delta x < 2\xi/\sqrt{3}$. a**, $U(x)$ due to two defects, each contributing a Lorentzian potential with $U_0 = 2$ eV and $\xi = 46.4$ nm for four values of separation $\Delta x$ between the defects, ranging from 0 nm to 52.5 nm. With increasing $\Delta x$, the potential gradually changes into a U shape, resulting in a reduction of the spring constant in the center of the well. The solid lines end at the inflection points of the potential. **b**, The corresponding restoring force $F_{dc} = \partial U/\partial x$ showing the development of an inflection point in the center of the well. **c**, Corresponding ac displacement $x_{ac}$ due to $F_{ac} = 16.5$ fN showing the development of large softening in the center of the well. **d**, Dependence of $x_{ac}$ at the softening point in the center of the well as a function of the separation $\Delta x$ between the defects. The softening diverges on approaching $\Delta x = 2\xi/\sqrt{3} = 53.6$ nm for the case of a Lorentzian potential.

## 9. Rate of thermally-activated vortex jumps

Vortices in a thin film of thickness d ~ ξ can be modeled as over-damped particles, so we evaluate the rate Γ of spontaneous hopping of vortices over the energy barrier ΔU between neighboring pins, using the standard theory of thermally-activated escape [46]:

$$\Gamma = \omega_0 e^{-\Delta U/kT}, \qquad \omega_0 = \frac{\omega_a \omega_b}{2\pi\gamma} \qquad (1)$$

where $\omega_a$ and $\omega_b$ are the frequencies of small oscillations at the bottom of the potential well and at the top of the inverted barrier, $\gamma = \eta/M$, M is the vortex mass per unit length, $\eta = \phi_0 B_{c2}/\rho_n$ is the Bardeen-Stephen vortex viscosity, $B_{c2} = \phi_0/2\pi\xi^2$ is the upper critical field and $\phi_0$ is the flux quantum, and $\rho_n$ is the normal state resistivity. We first estimate $\omega_a$ and $\omega_b$ for a vortex in a 1D tilted washboard potential U(x) = [1 − cos(2πx/s)]$U_0$ − J$\phi_0$x where J is the sheet current density. The vortex position x(t) is described by the sine-Gordon equation

$$dM\ddot{x} + d\eta\dot{x} + \frac{2\pi U_0}{s}\sin\frac{2\pi x}{s} = J\phi_0 \qquad (2)$$

where $U_0$ is the maximum pinning energy, s = Cξ is the characteristic spatial scale of the pinning site, and C ~ 1. Hence,

$$\omega_a^2 = \omega_b^2 = \frac{4\pi^2 U_0}{dM\xi^2 C^2}\sqrt{1-\beta^2} \qquad (3)$$

Here β = J/$J_c$, and $J_c$ = 2π$U_0$/s$\phi_0$ is the sheet critical current density. Near the depinning transition, J ≈ $J_c$ and the expansions of the activation barrier ΔU(β) and the effective attempt frequency $\omega_0$ in the small parameter 1 - β << 1 yield

$$\Delta U = \frac{4\sqrt{2}}{3}U_0(1-\beta)^{3/2} \qquad (4)$$

$$\omega_0 = \frac{4\pi^2 U_0 R_s \sqrt{2(1-\beta)}}{\phi_0^2 C^2} \qquad (5)$$

where $R_s = \rho_n/d$ is the film sheet resistance in the normal state. For the parameters of our Pb film, $U_0$ = 2 eV, and $R_s$ = 0.21 Ω, we obtain ΔU/ $k_B T$ = 1.042×10$^4$(1-β)$^{3/2}$ at T = 4.2K, and $\omega_0$ = 8.1×10$^{11}$(1-β)$^{1/2}$/C$^2$ Hz. We now estimate the critical value of $\beta_c$ at which thermal fluctuations cause spontaneous hopping of vortices between pinning sites within the time window $t_0$ ≅ 300 s of our measurements. Defining $\beta_c$ by the condition Γ$t_0$ ≈ 1, we have the following relationship

$$\frac{4\pi^2 U_0 t_0 R_s \sqrt{2(1-\beta_c)}}{\phi_0^2 C^2} exp\left[-\frac{4\sqrt{2}U_0}{3kT}(1-\beta_c)^{3/2}\right] = 1 \qquad (6)$$

For the materials parameters given above, and with C = 1, Eq. (6) yields 1 - $\beta_c$ ≅ 0.022, $\omega_0$ ≅ 1.2×10$^{11}$ Hz, and ΔU($\beta_c$) ≅ 34 $k_B T$ = 12 meV. As a result, the pinning spring constant α(β) = $\alpha_0$(1 - $\beta_c^2$)$^{1/2}$ ≈ $\alpha_0$/5 softens at the transition point β = $\beta_c$ by a factor of ~5 relative to the spring constant $\alpha_0$ at zero current.

The thermally-activated hopping rate can be enhanced by quantum effects even well above the crossover temperature $T_0$ = ħ$\omega_0$/k at which Γ becomes limited by the quantum tunneling of vortices [46]. For the above values, the quantum crossover temperature $T_0$ = 0.9 K at C = 1, well below T = 4.2 K. In this case Γ in Eq. (1) is increased by the factor q given by [46]

$$q = (\gamma^2 T_0/\omega_b^2 T)^{2T_0/T} \qquad (7)$$

For T = 4T$_0$, the factor $q$ = γ/2ω$_b$ = 0.5(ℏη/2πk$_B$T$_0$M)$^{1/2}$ can be large, but it depends on uncertain parameters such as the vortex mass which is determined by multiple mechanisms [47-49]. As an illustration, we evaluate an upper limit of $q$ taking into account only the contribution of localized quasiparticles in the vortex core to the mass, M = 2mk$_F$/π$^3$ where m is the effective electron mass and k$_F$ is the Fermi wave number [50]. Hence, q = (π/4)[ℏϕ$_0$B$_{c2}$/R$_n$dmk$_F$k$_B$T$_0$]$^{1/2}$. For k$_F$ = 10$^{10}$ m$^{-1}$, B$_{c2}$ = 153 mT and the parameters used above, we get q ≈ 3, which would give a negligible contribution to 1 - β$_c$ for the huge values of t$_0$Γ$_0$ and U$_0$/k$_B$T evaluated above. Even the much larger quantum factor q ~ 10$^3$ would increase the critical value of 1 - β$_c$ = 0.022 by only 17%.

## 10. Broken spring phenomenon

At the inflection point of the potential well the restoring force reaches its maximum and the spring constant $k = \partial^2 U/dx^2$ vanishes, resulting in diverging softening and infinite $x_{ac}$. On approaching the critical force the divergence is cut off by thermal fluctuations when the activation barrier $\Delta U(\beta_c) \cong 34$ k$_B$T for our experimental situation (see section S9). Figure 2a shows that for a potential well due to a single defect with $U_0$ of a few eV, $x_{ac}$ increases gradually towards the well edges and reaches a value that is about five times larger than at the bottom of the well when $\Delta U(\beta_c) \cong 34$k$_B$T. Correspondingly, the restoring force flattens towards its extrema points (Fig. 2c). In our experiment, this softening is much milder and $x_{ac}$ increases usually by less than a factor of two (Fig. 2e). The corresponding restoring force increases rather linearly with displacement and then vanishes quite abruptly with very little rounding, resembling the breaking of a spring (Fig. 2g). This behavior is inconsistent with a potential due to a single defect.

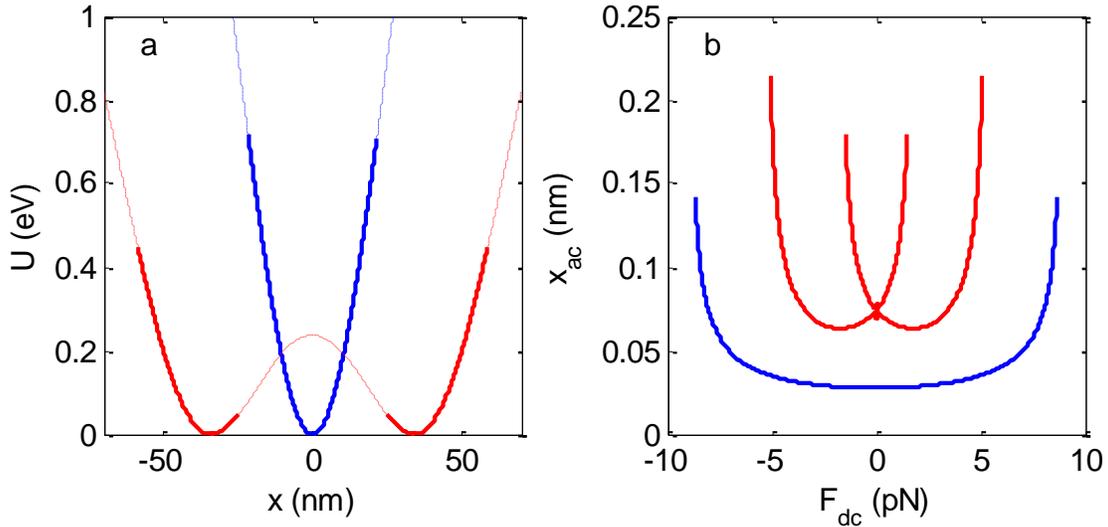

**Figure S10. Potential structure and vortex response in the case of two defects. a,** $U(x)$ due to two defects with $U_0 = 2$ eV when the two defects overlap and form a single well (blue) and when the defects are separated by $\Delta x = 80$ nm and form two nearby wells separated by a barrier comparable to our experimental findings (red). **b,** Vortex ac displacement $x_{ac}$ due to $F_{ac} = 16.5$ fN for the two cases showing similar softening near all the well edges and the absence of the broken-spring effect. The data points in **(a)** and **(b)** end where the activation barrier $\Delta U(\beta) = 34$ k$_B$T.

Figure S10 shows that two defects alone also cannot result in a broken-spring effect. In this case, when the defects are closer than $2\xi/\sqrt{3}$ a single well is formed with large softening in the center of the well (Figure S9), but the response near the edges of the well displays softening similar to that of the single defect case (not shown in Figure S9c for clarity). When the defects are separated by more than $2\xi/\sqrt{3}$, two separate wells are formed with a barrier between them. The outer edges of the well show softening as in the case of a single defect, while the softening on the inner sides of the two wells depends on the size of the barrier. For small barriers, thermal activation clearly leads to breaking of the spring between the two wells. However, this is not observed experimentally since the breaking of the spring is observed for large untilted barriers. When the distance between the two defects is increased to form a barrier that is comparable to experimental situation, no substantial spring breaking is attained (Figure S10). Thus, the presence of two nearby defects also cannot explain the observed broken-spring effect.

More generally, it can be shown that any combination of defects in 1D cannot lead to the broken-spring effect since the resulting $U(x)$ is always smooth on the scale of $\xi$. Surprisingly, we find that the situation is fundamentally different in 2D. A random distribution of defects in 2D still results in $U(x,y)$ that is smooth on the scale of $\xi$, however the vortex does not move along a straight line in this potential but rather forms nontrivial trajectories. The projection of the potential along this trajectory onto the $x$ axis (direction of the driving force) now results in $U(x)$ that may contain ripples on characteristic length scales significantly shorter than $\xi$, leading to the broken-spring effect as described in the main text. By performing numerical studies we find that the necessary condition for spring-breaking is that at least two nearby defects form a single central well that is broader than $\xi$ and that the additional adjacent defects create ripples in this 2D potential. Three defects are sufficient for breaking the spring on one side of the well, while four or more defects are required for spring-breaking on both sides of the well (Figure 5).

## 11. Vortex – SOT interaction

We can estimate the interaction force between the tip and the vortex as follows. In our geometry, we can approximate the SOT loop as a single segment of length $\Delta L = 177$ nm (equal to the SOT diameter) carrying current parallel to the sample. The current that flows in the segment is the current applied to the SOT, $I_{SOT} \cong I_c^{SOT} \cong 100$ µA. The current through the segment is applied through two parallel wires perpendicular to the sample surface and carrying current in opposite directions; these are the leads to the SOT. These parallel leads, perpendicular to the sample plane, do not experience a net force in the vortex field, whereas the horizontal segment experiences an in-plane interaction force of $F_{int} = I_{SOT} \Delta L B_z$, where $B_z \cong 2$ mT is the typical field created by the vortex at the SOT location, as measured directly by the SOT. With the above parameters we obtain $F_{int} \cong 35$ fN. This dc force between the tip and the vortex is about two orders of magnitude smaller than the typical restoring force $F_{dc}$ of a potential well and therefore should not affect our measurements. Although the interaction force is comparable to our typical $F_{ac} = 93.1$ fN, since it is dc, it should not influence the ac measurements.

In order to further test the possible influence of the tip, we have compared measurements of vortex hopping between two wells with and without scanning the SOT over the vortex. In the latter case, the SOT was stationary located away from both the wells, but not too far from the final well. When the vortex is located in the original well, it creates a very small $B_z$ at the location of the SOT and hence no interaction force. When the vortex arrives at its final well, a small $B_z$ increase is detected by the SOT. As a result, the hopping event can be observed without influencing the vortex at its original location. By sweeping $F_{dc}$, we have thus compared the hopping process with and without SOT scanning. The results showed no observable difference between the two procedures.

## 12. Numerical simulations

We generated various sets of randomly distributed defects in an area of 1.2×1.2 μm² to simulate the vortex dynamics. Each defect contributes a potential of the form $U_n(x,y) = \frac{-U_0}{1+\left(\frac{x-x_n}{\xi}\right)^2+\left(\frac{y-y_n}{\xi}\right)^2}$, with the total potential given by a linear superposition from all the n defects. The simulations in Figure 4 were performed using $U_0 = 0.66$ eV, a defect density of 200 μm⁻², and $\xi = 46.4$ nm. For each increment of $F_{dc}$ we trace the vortex flow by solving the massless dynamic equation of vortex velocity vector $\boldsymbol{v}$ given by $\boldsymbol{v} = -\nabla U$, where $U$ is the tilted potential. The viscous vortex drag is not taken into account since, even though the velocity is inversely proportional to the vortex viscosity, the resulting vortex trajectory is independent of the viscosity. As long as the vortex resides inside the well, the trajectory follows the minimum of the titled potential at each $F_{dc}$. When $F_{dc}$ exceeds the maximum restoring force of a well, the vortex flows along an intricate dynamic trajectory defined by the gradient of the tilted potential until it reaches the next minimum point (dashed lines in Figure 4c). Note that the color scale in Figure 4c represents the untilted potential. Similar to the experimental case, we perform subloops of the driving force in order to explore the full set of trajectories.

We have also explored additional functional forms of $U(x,y)$ of a defect, including a potential similar to that derived using Ginzburg-Landau calculations in ref [33]. The qualitative features described in the paper could be reproduced with the different defect potentials for comparable values of $U_0$ and $\xi$.

### 13. Ginzburg-Landau simulations of strong pinning

We solve the problem of depinning of a single vortex in a film using the static GL equation

$$\Delta\psi + (1 - p(x,y) - |\psi|^2)\psi = 0. \qquad (1)$$

Equation (1) is written in a dimensionless form, where all distances are measured in $\xi(T)$ and the order parameter $\psi(x,y)$ is in units of the equilibrium order parameter $\psi_0$. The magnetic vector potential responsible for the London screening near the vortex core is neglected. The function $p(x,y)$ quantifies the local depression of $T_c(x,y)$ that produces the pinning potential well. We used the following model form of $p(x,y)$:

$$p(x,y) = p_0(\tanh(a-r) + \tanh(a+r)), \qquad (2)$$

where $p_0$ and $a$ define the magnitude and the width of the well, respectively, $r = \sqrt{x^2+y^2}$, and the center of the pin is placed at the origin. Unlike the sharp rectangular well for a dielectric precipitate or a pinhole, $p(r)$ in Equation (2) has smooth tails that decay over the coherence length $\xi$, consistent with the static GL solution for the superconducting-normal domain well. This form of $p(r)$ models a proximity-coupled normal region, for example a circular normal island on the film surface, which suppresses superconductivity all the way across the film thickness $d \cong \xi$.

We solve Equations (1) and (2) on a $20 \times 20 \, \xi^2$ area, which is large enough to prevent the boundaries from influencing the depinning critical current. An external dc current is imposed in the $y$ direction and a zero perpendicular current condition is imposed across the side boundaries. The solution of this problem gives the dependence of the vortex displacement $u(J)$ as a function of the dc current density $J$ normalized to the critical current density $J_c$ (Figure S11). Here, the vortex core coordinate $u(J)$ is defined as the position of the topological singularity in the phase of the order parameter. The dependence $u(J)$ terminates at $J = J_c$ so that the GL equations do not have static vortex solutions at $J > J_c$. The results of this simulation for different values of the potential depth $p_0$ and $a = 1.5$ are shown in Figure S11. The displacement $u(J)$ monotonically increases with $J$ and the pinning spring constant softens noticeably as $J$ approaches $J_c$. The overall behaviors of the restoring force and the potential are similar to that of the model Lorentzian $U(r)$ considered in the main text.

The full GL calculation thus does not produce the broken-spring behavior for a single defect even in the case of strong pinning, so the concept of rigid pinning wells can be used, at least qualitatively, for both weak and strong pinning. The latter is important because pinning energies in Pb films suggest strong pinning with $p_0 > 1$.

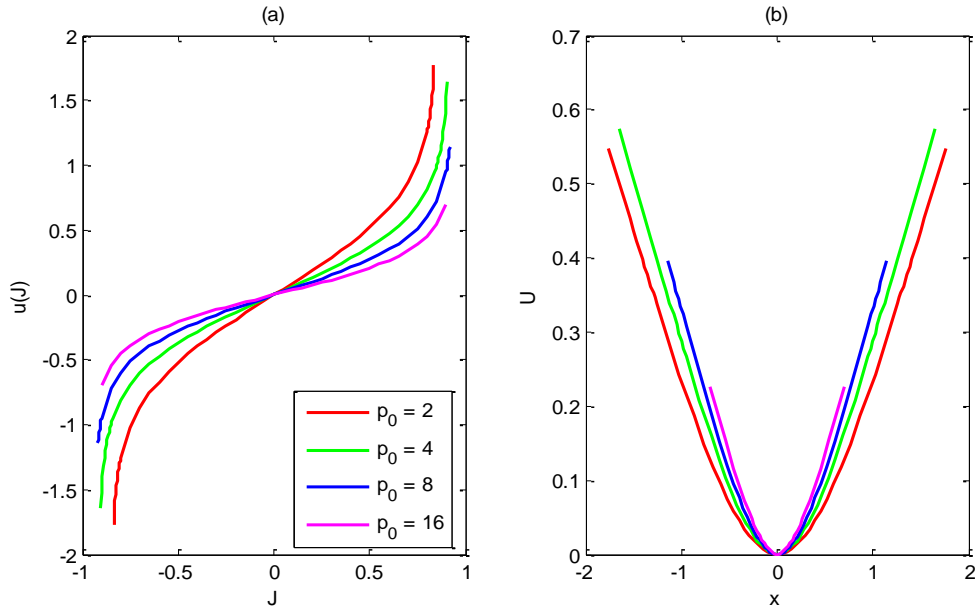

**Figure S11. GL calculation of a strong pinning potential. a,** The vortex displacement $u(J)$ as a function of the dc current density $J$ for different values of $p_0$ and $a = 1.5$. **b,** Reconstruction of potential wells for different values of $p_0$ from the displacement plot **(a)**.

**References:**

46. Weiss, U. *Quantum Dissipative Systems, 4-th Ed.* World Scientific Publishing Co., Singapore (2012).
47. Kopnin, N. B. & Vinokur, V. M. Dynamic vortex mass in clean superconductors and superfluids. *Phys. Rev. Lett.* **81**, 3952-3955 (1998)
48. Sonin, E. B., Geshkenbein, V. B., van Oterlo, A. & Blatter, G. Vortex motion in charged and neutral superfluids: A hydrodynamic approach. *Phys. Rev.* B **57**, 578-581 (1998)
49. Chudnovsky, E. M. & Kuklov, A. B. Inertial mass of Abrikosov vortex. *Phys. Rev. Lett.* **91**, 067004 (2003).
50. Suhl, H. Inertial mass of a moving fluxiod. *Phys. Rev. Lett.* **14**, 226-230 (1965).